\documentclass[useAMS,usenatbib]{mn2e}
\usepackage{graphicx}

\newcommand{\kms}{km\,s$^{-1}$}     
\newcommand{\sqcm}{cm$^{-2}$}  
\newcommand{\lya}{Ly$\alpha$}
\newcommand{\hi}{\mbox{H\,{\sc i}}}

\newcommand{\os}{\mbox{O\,{\sc vi}}}
\newcommand{\cf}{\mbox{C\,{\sc iv}}}
\newcommand{\ct}{\mbox{C\,{\sc iii}}}
\newcommand{\sif}{\mbox{Si\,{\sc iv}}}
\newcommand{\nf}{\mbox{N\,{\sc v}}}
\newcommand{\mgii}{\mbox{Mg\,{\sc ii}}}
\newcommand{\nsys}{35}   % Num O VI systems %*
\newcommand{\ncomp}{101} % Num O VI comp
\newcommand{\Nstr}{Nine} % Num Strong
\newcommand{\nstr}{9}    % Num Strong
\newcommand{\nwea}{26}   % Num Weak %*
\newcommand{\zabs}{$z_{\rm abs}$}
\newcommand{\zqso}{$z_{\rm qso}$}

\title[High-$z$ Proximate \os]
      {A Study of Quasar Proximity in \os\ absorbers at
      $z$=2--3\thanks{Based on observations taken under ESO
      Programme ID 166.A-0106(A), using the Ultraviolet and Visual Echelle
      Spectrograph (UVES) on the Very Large Telescope, Unit 2 at
      Paranal, Chile.}} 
\author[Fox, Bergeron, \& Petitjean]
       {Andrew J. Fox$^{1,2}$\thanks{E-mail: afox@eso.org}, 
	 Jacqueline Bergeron$^2$, and Patrick Petitjean$^{2,3}$\\
$^1$European Southern Observatory, Alonso de C\'ordova 3107, Casilla
      19001, Vitacura, Santiago 19, Chile\\
$^2$Institut d'Astrophysique de Paris, UMR 7095 CNRS,
  Universit\'e Pierre et Marie Curie, 98bis Boulevard Arago, 75014
  Paris, France\\
$^3$LERMA, Observatoire de Paris, 61 Avenue de l'Observatoire, 75014
      Paris, France} 

\begin{document}
\maketitle

\begin{abstract}
With the goal of investigating the nature of \os\ absorbers at high
redshifts, we study the effects of proximity to the background quasar.
In a sample of sixteen quasars at \zqso\ between 2.14 and 2.87
observed at high signal-to-noise and 6.6~\kms\ resolution with VLT/UVES,
we detect \nsys\ \os\ absorption-line systems (comprising over 100
individual \os\ components) lying within 8\,000~\kms\ of \zqso.
We present component fits to the \os\ absorption and the accompanying
\hi, \cf, and \nf. The systems can be categorized into \nstr\ strong
and \nwea\ weak \os\ absorbers. 
The strong (intrinsic) absorbers 
are defined by the presence of either broad, fully saturated %mini-BAL
\os\ absorption or partial coverage of the continuum source, and in
practice all have log\,$N$(\os)$\ga$15.0; these systems are interpreted as
representing either QSO-driven outflows or gas close the central engine
of the AGN. 
The weak (also known as narrow) systems show no partial coverage or
saturation, and are characterized by log\,$N$(\os)$<$14.5 and a median
total velocity width of only 42~\kms. %*  
The incidence d$N$/d$z$ of weak \os\ systems within 2\,000~\kms\ of
the quasar down to a limiting equivalent width of 8~m\AA\ is %*
42$\pm$12. Between 2\,000 and 8\,000~\kms, d$N$/d$z$ falls to 14$\pm$4, %**
equal to the incidence of intervening \os\ absorbers measured in
the same spectra. 
Whereas the accompanying \hi\ and \cf\ column densities are
significantly lower (by a mean of $\sim$1~dex) %*
in the weak \os\ absorbers within 2\,000~\kms\ of \zqso\
than in those at larger velocities, %*
the \os\ column densities display no dependence on proximity.
Furthermore, significant offsets between the \hi\ and \os\ centroids
in $\approx$50 per cent of the weak absorbers imply that (at least in
these cases) the \hi\ and \os\ lines are not formed in the same phase of
gas, preventing us from making reliable metallicity determinations,
and ruling out single-phase photoionization-model solutions.
In summary, we find no firm evidence that quasar radiation influences
the weak \os\ absorbers, suggesting they are collisionally ionized rather than
photoionized, possibly in the multi-phase halos of foreground galaxies.  
Non-equilibrium collisional ionization models are needed to explain
the low temperatures in the absorbing gas, which are implied by narrow
line widths ($b\!<\!14$~\kms) in over half of the observed \os\
components. %* non-solar abundances
\end{abstract}


\section{Introduction}
Quasars are ideal backlights for absorption-line spectroscopy, thanks
to their high luminosities and flat continua. 
Yet in addition to allowing us to detect foreground absorbers, quasars
also create and influence them, through the effects of outflows and
ionizing radiation. The enhanced level of ionizing radiation in the
vicinity of quasars gives rise to the line-of-sight proximity effect,
the observed decrease in the number density of
\lya\ forest lines (equivalent to a decrease in the mean optical
depth, and an increase in the level of hydrogen ionization) at velocities
approaching the quasar \citep{Ca82, Mu86, Ty87, Ba88, Lu91, Bh94,
  Ra98, Sc00, Gu07, FG08}.

To obtain an unbiased view of absorbing structures in the high-$z$
Universe, one needs to remove the effects of proximity from samples of
metal-line systems.
The broad absorption line (BAL) systems, easy to identify in QSO
spectra because of their optically thick absorption extending over
thousands of \kms, are clearly connected to the
quasar. BALs trace high-velocity QSO outflows at speeds of up to 0.1$c$ 
\citep[e.g.][]{Tu84, Tu88}, and are seen in 10--15\% of quasar
spectra at $z\!>\!1.5$ \citep{We91, Tr06}.
A separate category of mini-BAL systems, with fully saturated absorption
profiles and total velocity width $<$ 2\,000~\kms, has also been
identified \citep{Tu88, Ch99, Yu02, Mi07b}.
As with BALs, mini-BALs are interpreted as tracing outflows ejected by QSOs.
A more difficult task is to decide whether {\it narrow}
absorbers in quasar spectra are either created by or ionized by the
QSO. The traditional way to do this employs the displacement velocity
from the quasar,   
$\delta v\!\equiv\!v_{\rm qso}\!-\!v_{\rm abs}$.
So-called `associated' or `$z_{\rm abs}\!\approx\! z_{\rm qso}$' or
`proximate' absorbers, typically defined as narrow absorbers at 
$\delta v\!<\!5\,000$~\kms\ \citep{We79, Fo86, An87, HF99} 
are often removed from intervening
samples because of the possibility they may be intrinsic\footnote{We
  favour the use of the term `proximate' since it involves no 
  assumption about the absorber origin.}. 
Arguments in favour of an intrinsic nature for many proximate
absorbers include time-variability of absorption,
partial coverage of the continuum source, 
detection of excited ionic states implying high gas densities, 
profiles that are smoother than intervening absorbers,
and super-solar metallicities 
\citep{Wa93, Mo94, Pe94, Sv94, Tr96, BS97, Ha97,
  Ha97a, Ha97c, Ha01, Ha00, PS99, SP00, Ga99, Ga01, Ga06, Na04}.

Using a velocity cutoff to differentiate intervening and
intrinsic absorbers has two main problems:
quasar-ejected intrinsic systems can appear at higher $\delta v$
\citep{Ha97b, Ri99, Ri01, Mi07a, Ne08}, and intervening systems
can appear at lower $\delta v$ \citep{Mo98, Se04}. 
So are many genuine intervening systems lost when they are excluded 
from absorber samples just because of their proximity to the quasar?
In this paper we investigate the transition between proximate and
intervening absorbers. We focus on \os\ absorbers, which
trace either warm-hot ($T\!\ga\!10^5$~K)
collisionally-ionized plasma, or photoionized gas
subject to a hard ionizing spectrum extending to energies above
113.9~eV (the ionization potential to create O$^{+5}$).  
\os\ absorbers are of interest 
for many reasons, including the significant role they play in the
baryon and metal budgets, the window they provide on intergalactic
metal enrichment, and their ability to trace energetic galaxy/IGM
interactions such as accretion and galactic winds. After the first
detection of \os\ absorbers at $z$=2--3 by \citet{LS93}, much work has
been done in the era of 10m-class telescopes to characterize their properties
\citep*{Sy00, Ca02, Be02, Si02, Si04, Si06, Be05, Lo07, Ag08, Go08}. 

Here we ask a simple question: is there a proximity effect in \os?
In other words, can we see the signature of quasar photoionization, in
terms of correlations between the ionization properties of absorbers
and their proximity to the quasar? 
To address this issue, we form a sample of
proximate \os\ absorption systems at $z$=2--3 using a homogeneous set
of high-resolution VLT/UVES quasar spectra, and we then compare their
properties to a sample of intervening (i.e., non-proximate) \os\
absorbers observed in the same set of spectra; 
the intervening sample has been published by \citet[][hereafter BH05]{Be05},
but it has since been enlarged from ten to twelve sight lines. %*
We also briefly compare our results on \os\ absorbers at $z$=2--3 to those
obtained in the low-redshift Universe with space-based ultraviolet
spectrographs. This paper is thus a study of both the proximity effect and 
the nature of \os\ absorbers in general, and is
structured as follows. In \S2 we describe the data
acquisition and reduction, and our absorber identification and
measurement processes. In \S3 we discuss the observational
properties of proximate \os\ absorbers. In \S4 we discuss the
implications of our results, and we then present a summary in \S5.
Throughout this paper we adopt a WMAP 3-year cosmology with
$\Omega_{\rm M}$=0.27, $\Omega_\Lambda$=0.73, and $h_{\rm 70}$=1 \citep{Sp07}. 

\section{Data Acquisition and Handling}
\subsection{Observations}
The ESO-VLT Large Programme {\it The Cosmic Evolution of the
Intergalactic Medium} (IGM) has built a homogeneous sample of
high resolution, high signal-to-noise quasar spectra for use in
studying the high-redshift IGM. 
Twenty bright quasars (most with $V\!<\!17$) in the redshift range
2.1--3.8 (median \zqso=2.44) were selected for observation, 
deliberately avoiding sight lines that contain high-column density damped
\lya\ absorbers. The observations were carried out with the
Ultraviolet and Visual Echelle Spectrograph 
\citep[UVES;][]{De00} on VLT UT2 at Paranal, Chile, using a
total of thirty nights of service-mode telescope time 
between June 2001 and September 2002.
The observations were taken in good seeing conditions ($<$0.8\arcsec)
and at airmass$<$1.4, with both blue and red dichroics 
(Dic1 346+580, Dic1 390+564, and Dic2 437+860 settings),
giving spectra complete from 3\,000
to 9\,000~\AA. The data reduction was conducted using
the {\sc MIDAS} pipeline described in \citet{Ba00}, and then continua
were fitted using a third-degree spline function interpolated between
regions free from absorption. Full details of the reduction process
are given in \citet{Ch04} and \citet{Ar04}. 
The 2$\times$2 binning mode was used, giving a rebinned
pixel size of 2.0--2.4~\kms. 
The spectral resolution of the data, $R\approx45\,000$, corresponds to
a velocity resolution of $\approx$6.6~\kms\ (FWHM). Typical
signal-to-noise ratios per pixel in the spectra are $\approx$35 at
3\,500~\AA\ and $\approx$60 at 6\,000~\AA.
Among the twenty quasars, sixteen are at $z\!<\!3$. 
These form the basic sample for this paper. The four cases at
$z\!>\!3$ were excluded, because the density of the \lya\ forest at these
redshifts renders futile searches for \os.

The emission-line redshifts of ten of the sixteen quasars 
in our sample have been carefully determined by \citet[][and
  references therein]{Ro05}, 
and we adopt their values. For the remaining six quasars we take
\zqso\ from \citet{Sc06}. Each of these quasar redshifts is based on 
a combination of the observed H$\alpha$, \mgii, \cf, and \sif\
emission-line redshifts, allowing for systematic shifts in the
displacement velocities of the various lines
\citep[e.g.][]{TF92}. The uncertainty on each \zqso\ is of the order
500~\kms\ \citep{Ro05}. 

\subsection{Identification of \os\ systems}
We searched systematically through the sixteen quasar spectra in our sample
for \os\ absorbers within 8\,000~\kms\ of the quasar redshift.
This velocity range corresponds to redshifts between 
$z_{\rm min}$=\zqso--[(1+\zqso)\,8\,000~\kms/c]
and \zqso. 
Our choice of velocities up to 8\,000~\kms\ from the quasar was
made in order to investigate whether a transition occurs at
5\,000~\kms, the conventional cutoff between proximate and
intervening systems. In practise, we also searched for
(and found) \os\ absorbers at up to 2\,500~\kms\ {\it beyond} 
the quasar redshift, indicating the presence of high-velocity inflows
toward the quasar.

32 \os\ absorbers were identified by their presence in both lines of the
\os\ $\lambda\lambda$1031.926, 1037.617 doublet in the correct ratio
through the line profile, i.e.  
$\tau_a(v, \lambda 1031)/\tau_a(v, \lambda 1037)=2$, 
where the apparent optical depth
$\tau_a(v)={\rm ln}\,[F_c(v)/F(v)]$, and where $F(v)$ and $F_c(v)$ are the
observed flux level and estimated continuum flux level, respectively,
as a function of velocity.
In addition, three \os\ absorbers were identified in one line of the
\os\ doublet only, with partial blending in the other \os\ line, but
with corresponding absorption seen in \cf\ and \hi\
at exactly the same redshift. %*
Together the sample consists of the \nsys\ \os\ systems listed in Table 1.
We define the system redshift \zabs\ by the position of the
strongest \os\ component in the system, i.e. the wavelength of maximum
optical depth. In each system, we looked for accompanying absorption
in \hi\ (Ly$\alpha$, Ly$\beta$, and Ly$\gamma$), \ct\ $\lambda977$,
\cf\ $\lambda\lambda1548,1550$, and \nf\ $\lambda\lambda1238,1242$ at
the same redshift \zabs. 

We denote the velocity displacement of each absorber from the quasar as 
$\delta v\equiv c|z_{\rm qso}\!-\!z_{\rm abs}|/(1\!+\!z_{\rm abs})$.
%For simplicity, we refer to $\delta v$ as the proximity. 
A negative value corresponds to an absorber moving toward the quasar.
The \os\ systems almost always contain multiple components clustered 
together. We thus define $\delta v$ both for each absorber
and for each individual \os\ component.
The uncertainties on all velocity displacements are dominated by the
$\approx$500~\kms\ uncertainty of the quasar emission redshifts.

\subsection{Classification of \os\ systems}
After identifying \nsys\ proximate \os\ systems, we 
found that they could be classified into:

(a) Seven {\bf strong} \os\ systems, which are defined by showing 
both fully saturated \os\ lines extending over tens to hundreds of
\kms, and strong (often saturated) \nf\ and \cf\ absorption.
In practice we find that these seven systems all show
log\,$N$(\os)$>$15.0 and are at $\delta v\!<\!3\,000$~\kms, with four
of the seven at $\delta v<0$ (\zabs$>$\zqso), i.e. the strong
systems are strongly clustered around \zqso. %*
The strong systems are interpreted as being {\bf intrinsic} to the
quasar. Three of the strong systems are classified as
mini-BALs\footnote{The mini-BALs are the absorbers at \zabs=2.4426
  toward HE~1158-1843, 2.1169 toward HE~1341-1020, and \zabs=2.9041
  toward HE2347-4342.}, but no full (classical) BAL systems are
present in our sample. Partial coverage of the continuum source is
present in many of the strong systems, as evidenced by the failure
to find a successful Voigt profile fit to the \os\ profiles, even
though the velocity structure is common among the two members of the
doublet. This confirms an intrinsic origin, because partial coverage
of the continuum source only occurs when the absorbing gas is physically
close to the AGN. The strong systems contain multiple components, with
a mean of 5.9 components per system. %* 

(b) \nwea\ {\bf weak} \os\ systems, also known as {\bf narrow} systems,
which are unsaturated and show no evidence for partial
coverage, i.e. we were able to conduct 
successful Voigt profile fits to the two lines of the \os\ doublet.
The weak systems all have 12.80$<$log\,$N$(\os)$<$14.50, clearly
separated from the strong systems, but coincident with the 
range of \os\ column densities
observed in intervening systems (BH05). %*
Their total line widths are narrow, with a median and standard
deviation of $\Delta v_{90}$=42$\pm$33~\kms, %* 
where $\Delta v_{90}$ is the line width encompassing the central 90\%
of the integrated optical depth.
The weak \os\ systems are accompanied by \cf\
absorption in 24/26 cases, and in 9/26 cases by \nf. %*
An mean of 1.9 components per system is required to
fit these weak absorbers.%*

There were two intermediate cases that presented difficulties in their
classification: the systems at $z$=2.6998 and 2.7134 toward HE~0151-4326.
Neither of these systems shows a saturated \os\ profile, and their
\os\ column densities lie in-between the values shown by the strong
($>$15.0) and weak ($<$14.5 populations). However, differences between
the profiles of the two doublet lines of \os, and between the two
lines of \nf, in similar velocity ranges are strongly suggestive of
partial coverage. This implies they should be classified as intrinsic.
For this reason we place these systems in the strong category, and our
final sample consists of \nstr\ strong systems and \nwea\ weak systems.

In addition to the different \os\ (and \nf\ and \cf) column densities,
there are several lines of evidence that support the notion that the
strong and weak samples trace physically different populations.
The strong population have velocities (relative to \zqso) that
distribute differently, with 4/9 at \zabs$>$\zqso, whereas 22/26 weak
absorbers are at \zabs$<$\zqso. The strong population are show
markedly different \hi/\os\ ratios than the weak population (see
\S3.5). Finally absorbers in the strong population frequently show
evidence for partial coverage of the continuum source.
Whereas the strong absorbers are clearly associated with the
quasar \citep[e.g.][]{Tu84, We91, HF99}, either tracing QSO outflows
or inflows near the central engine of the AGN, the
origin of the weak \os\ absorbers is less clear. They are often called
`associated' because of their proximity in velocity to the quasar, but
there is not necessarily any physical connection. 
%between the absorber and the AGN.
In the rest of this
paper we investigate the properties of this weak, proximate sample.

\subsection{Component fitting}
We used the {\sc VPFIT} software
package\footnote{Available at http://www.ast.cam.ac.uk/$\sim$rfc/vpfit.html.} 
to determine the column densities, line widths, and central velocities
of the \os\ components in each system. 
The number of components to be fit was usually
self-evident from inspection of the line profiles, but in case of doubt we
used the minimum number of components necessary.
Our sample of \nsys\ \os\ systems comprises \ncomp\ components.
We also fit, when possible, the \hi, \nf, \ct, and \cf\ absorption,
independently of the \os\ fits, i.e. we did not fix any component
velocity centroids or line widths when fitting the various ions.
In the case of the \hi\ lines, we simultaneously  
fit \lya\ with whichever of Ly$\beta$, Ly$\gamma$, and
Ly$\delta$ were unblended. 
The total column densities in each ion in each system were obtained by
summing the component column densities, with the individual column
density errors added in quadrature to produce the overall error estimate.
All the Voigt profile fits are included in Figure 1, which shows the
absorption-line spectra for each system in our sample.
The total column density of each high ion and of \hi\ in each system
is given in Table 1. 
The detailed results of the component fits to \os, \nf,
\cf, and \hi\ in the weak sample are given in Tables 2 and 3.
All velocities in Tables 2 and 3 are given on a scale relative to the
system redshift \zabs.

In addition to the component fits, we measured the total column
densities in each system using the apparent
optical depth (AOD) method of \citet{SS91}. This method has the
advantage of requiring no assumptions about the component structure:
it simply requires a choice of minimum and maximum velocities for the
integration. It will return accurate column densities so long as
saturated structure is not present. 
Whereas saturation is obvious if resolved, because the flux goes to zero,
unresolved saturation can be difficult to detect.
In the case of a non-detection of \cf\ or \nf, we measured
the (null) equivalent width in this ion over the velocity range
in which \os\ is detected, and found the
3$\sigma$ error on this measurement. We then converted this maximum
allowed equivalent width into a column density limit using a linear
curve-of-growth. For example, if the equivalent width measurement was
0$\pm$10~m\AA, we derived a 3$\sigma$ column density limit 
$N_{\rm lim}$ using ${\rm EW_{lim}}$=30~m\AA\ and the relation 
$N_{\rm lim}=1.13\times10^{17}{\rm EW_{lim}}/\lambda^2f$ 
(with $N_{\rm lim}$ in \sqcm\ and $\lambda$ in \AA). Atomic data were
  taken from \citet{Mo03}.

In many of the strong absorbers, partial coverage of the continuum
source is present. Indeed, we use this as a criterion to classify an
absorber as intrinsic. Partial coverage is indicated by a similar
velocity structure in the two members of a doublet, but with
inconsistent optical depths in each ion.
In the case of partial coverage the equation
$F(v)=F_0e^{-\tau(v)}$ is no longer valid \citep[e.g.][]{HF99}, and a
velocity-dependent coverage fraction must be introduced. For these
cases, the column density derived from an AOD integration should be
considered as a lower limit. 

\section{Results}
In this section we present a discussion of the overall properties of
the population of proximate \os\ absorbers at $z$=2--3. Although we
discuss ionization processes, a detailed discussion of the physical
conditions in the individual systems that make up our sample is beyond
the scope of this paper. Some of the systems in our sample are
discussed in more detail elsewhere. These include the proximate
absorbers toward HE~2347-4342 \citep{Fe04}, and toward HE~1158-1843,
PKS~0329-255, and Q0453-423 \citep{DO04}. 

\subsection{Location of absorbers in velocity space}
The number of proximate \os\ absorption line systems seen in each
quasar spectrum varies between zero and five.
There are two quasar spectra in our sample without \os\ absorption 
detected within 8\,000~\kms\ of \zqso:
HE~2217-2818 (\zqso=2.41) and Q0002-422 (\zqso=2.77).
There are two quasars with five proximate \os\ absorbers in their
spectra: PKS~0237-23 (\zqso=2.233) and HE~1341-1020 (\zqso=2.135).
To illustrate the location in velocity space of the absorbers, we show
in Figure 2 the velocity displacement of all absorbers, strong and weak, for
each quasar. Over the whole sample (16 spectra) we find a mean of 
0.6 strong proximate \os\ systems per sight line, and 
1.6 weak proximate \os\ absorbers per sight line. %*
However, the \os\ sample is incomplete, because of regions where
\os\ systems are missed due to blending by the Ly$\alpha$
forest. As we show in \S3.7, %*
when we calculate the corrected redshift path for the \os\ search, our
estimated \os\ completeness is 84\% down to a limiting equivalent
width of 8~m\AA. %*

As a useful gauge of the velocity interval in which we expect to see
proximity effects, we can derive the distance from each quasar at
which the flux of 
extragalactic background (EGB) radiation equals the flux of quasar
radiation, i.e. we can derive an estimate of the size of the quasar
sphere-of-influence. The monochromatic quasar ionizing luminosities
at the Lyman Limit
have been derived for ten of the sixteen quasars by \citet{Ro05}, 
who calculated $L_{912}$ by extrapolating the QSO 
$B$-magnitudes assuming a spectral slope of $\alpha$=$-$0.5 
(where $F_\lambda\!\sim\!\lambda^\alpha$).
The EGB flux at the Lyman Limit at $z$=2.5 has been calculated to be 
$F_{\rm EGB}\!\approx\!6\!\times\!10^{-21}$~erg~cm$^{-2}$~s$^{-1}$~Hz$^{-1}$
\citep[][corresponding to 
log\,$J_\nu$=$-$21.3~erg~cm$^{-2}$~s$^{-1}$~Hz$^{-1}$~sr$^{-1}$; 
the values for the $z$=2 and 3 cases are not significantly different]{HM96}.
The distance $D$ at which $F_{\rm QSO}$=$F_{\rm EGB}$ is given by
$D$=$\sqrt(L_{912}/4\pi F_{EGB}$).
In the case where the velocities are Hubble flow-dominated, we can
convert our calculated values of $D$ to velocity displacement from the
quasar, using a value for the Hubble parameter calculated at the
redshift of each 
quasar using $H^2(z)=H^2_0[\Omega_{\rm M}(1+z)^3+\Omega_\Lambda]$.
With our adopted cosmology, $H(2.5)$=246~\kms~Mpc$^{-1}$.
These results are shown in Table 4, and are included in Figure 2, %*
where it can be seen that if the velocity displacements from the QSO
are due to the Hubble flow, then the proximity effect should only
extend to $\sim$1\,000--2\,500~\kms. While peculiar velocities are to be
expected, this is still a useful order-of-magnitude estimate for the
size of the quasars' ionizing sphere-of-influence.

\setcounter{figure}{1}
\begin{figure}
\includegraphics[width=9.2cm]{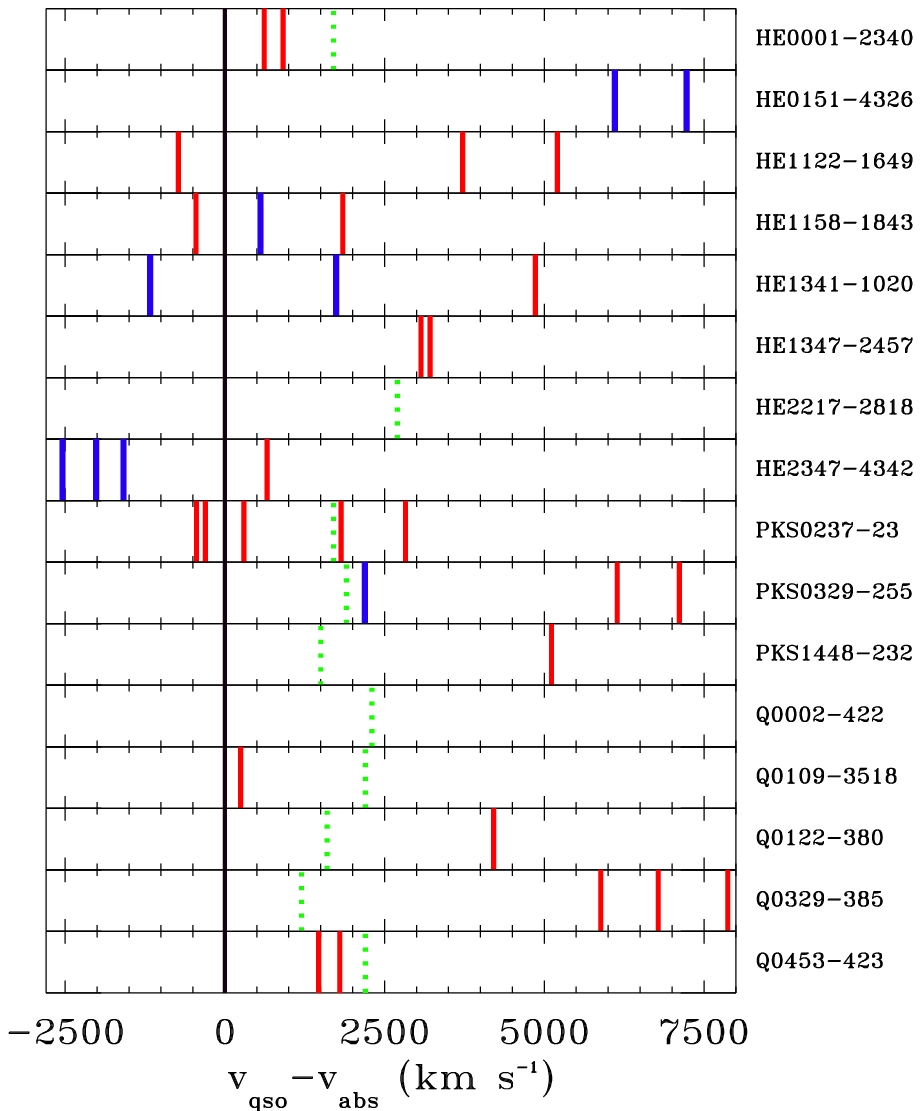}
\caption{Graphical display of the location in velocity space of the
  \os\ absorbers in this study, relative to their background quasar.
  Weak absorbers are shown in red, and strong absorbers in blue.
  The location of $\delta v$=0, i.e. \zabs=\zqso, is
  shown with a solid black line. Negative values for 
  $v_{\rm qso}\!-\!v_{\rm abs}$ imply the absorber is approaching the
  quasar. Dashed green lines show the velocity at which 
  $F_{\rm QSO}$=$F_{\rm EGB}$ assuming Hubble flow (see text).} 
\end{figure}

\setcounter{table}{3}
\begin{table}
\begin{minipage}{\columnwidth}
\caption{Size of quasar spheres-of-influence at 912~\AA.}
\begin{tabular}{lccc}
\hline
QSO & log\,$L_{912}$\footnote{Monochromatic luminosity at Lyman Limit,
  derived from the $B$ magnitude and $F_\lambda\sim\lambda^{-0.5}$
  \citep{Ro05}.}  
& $D$\footnote{Distance at which $F_{\rm QSO}$=$F_{\rm EGB}$=
  $6\!\times\!10^{-21}$~erg~cm$^{-2}$~s$^{-1}$~Hz$^{-1}$ 
  at $z$=2.5 \citep{HM96}.} 
& $\delta v$\footnote{Velocity separation from QSO at $D$ assuming Hubble
  flow, with $H$($z$) calculated at \zqso. $H$($z$=2.5)=246~\kms~Mpc$^{-1}$.}\\
 & (erg~s$^{-1}$~Hz$^{-1}$) & (Mpc) & (\kms)\\
\hline
  HE~0001-2340 & 31.65 &  7.7 & 1700 \\ 
  HE~2217-2818 & 31.99 & 11.4 & 2600 \\ 
   PKS~0237-23 & 31.67 &  7.8 & 1700 \\ 
  PKS~0329-255 & 31.58 &  7.1 & 1800 \\ 
  PKS~1448-232 & 31.53 &  6.7 & 1400 \\ 
    Q~0002-422 & 31.72 &  8.3 & 2200 \\ 
   Q~0109-3518 & 31.82 &  9.3 & 2200 \\ 
    Q~0122-380 & 31.63 &  7.5 & 1600 \\ 
    Q~0329-385 & 31.28 &  5.0 & 1100 \\ 
    Q~0453-423 & 31.71 &  8.2 & 2100 \\ 
\hline
\end{tabular}
\end{minipage}
\end{table}

\subsection{Presence of other ions}
The presence and strength of absorption in other ionic species in the
proximate \os\ absorbers provides important information on the
ionization conditions. \hi\ absorption is detected in each of the
\nsys\ \os\ systems, always in \lya\ and occasionally up to Ly$\delta$. 
Higher-order Lyman lines may also be present, but they lie deep in the
\lya\ forest where blending and low signal-to-noise aggravate the
search for detections. All but two of the
\os\ systems show \cf\ detections (the exceptions are the absorbers at
\zabs=2.2378 toward PKS~0237-23 and \zabs=2.6610
toward PKS~0329-255, both of which are weak \os\ systems with low
\os\ column densities). 
\nstr/\nstr\ strong \os\ systems contain \nf, 
whereas 9/\nwea\ weak \os\ systems contain \nf. %*.
Therefore while the presence of \nf\ is a feature of 
all strong systems, it does not imply that an \os\ system is strong.
Finally, \ct\ absorption is seen in 13/\nwea\ weak systems, and 7/\nstr\ %*
strong systems. However, because \ct\ $\lambda$977 lies in the
\lya\ forest, it is often blended and hence not detectable
even if present, so these fractions should be treated as lower limits.

\subsection{\os\ system column densities}
A plot of the total \os\ column densities in each absorber versus
proximity (Figure 3, top panel) shows the distinction
between the strong and weak populations. 
The strong population all show log\,$N\!\ga\!15.0$
and are(with two exceptions) clustered around \zabs=\zqso, 
whereas the weak population lies between the
detection limit of log\,$N\!\approx\!12.80$
(corresponding to a limiting equivalent width of 8~m\AA) and 14.5. 
Once the strong systems have been removed, the underlying weak
population shows no dependence of $N$(\os) with proximity to the quasar.
To investigate this further, we split the sample into \os\ absorbers
at $\delta v\!<\!2\,000$~\kms\ and those at 
$2\,000\!<\!\delta v\!<\!8\,000$~\kms. This division is motivated by
an observed enhancement in d$N$/d$z$ below 2\,000~\kms\ (\S3.7), as
well as this representing the expected size of the proximity zone (\S3.1).
A two-sided Kolmogorov-Smirnov (K-S) test shows that the probability
that the distribution of log\,$N$(\os) for absorbers at $<$2\,000~\kms\ is
different from those at $>$2\,000~\kms\ is only 17\%. %*
There is also no clear correlation between the quasar redshift
and the total \os\ column, as shown in the lower panel of Figure 3. 
We do note that the ratio of weak systems to strong systems changes
from 6.3 (19/3) at $z\!<\!2.5$ to 1.2 (7/6) at $z\!>\!2.5$, but the
statistics are too small for us to investigate this further. %*

\begin{figure}
\includegraphics[width=9.2cm]{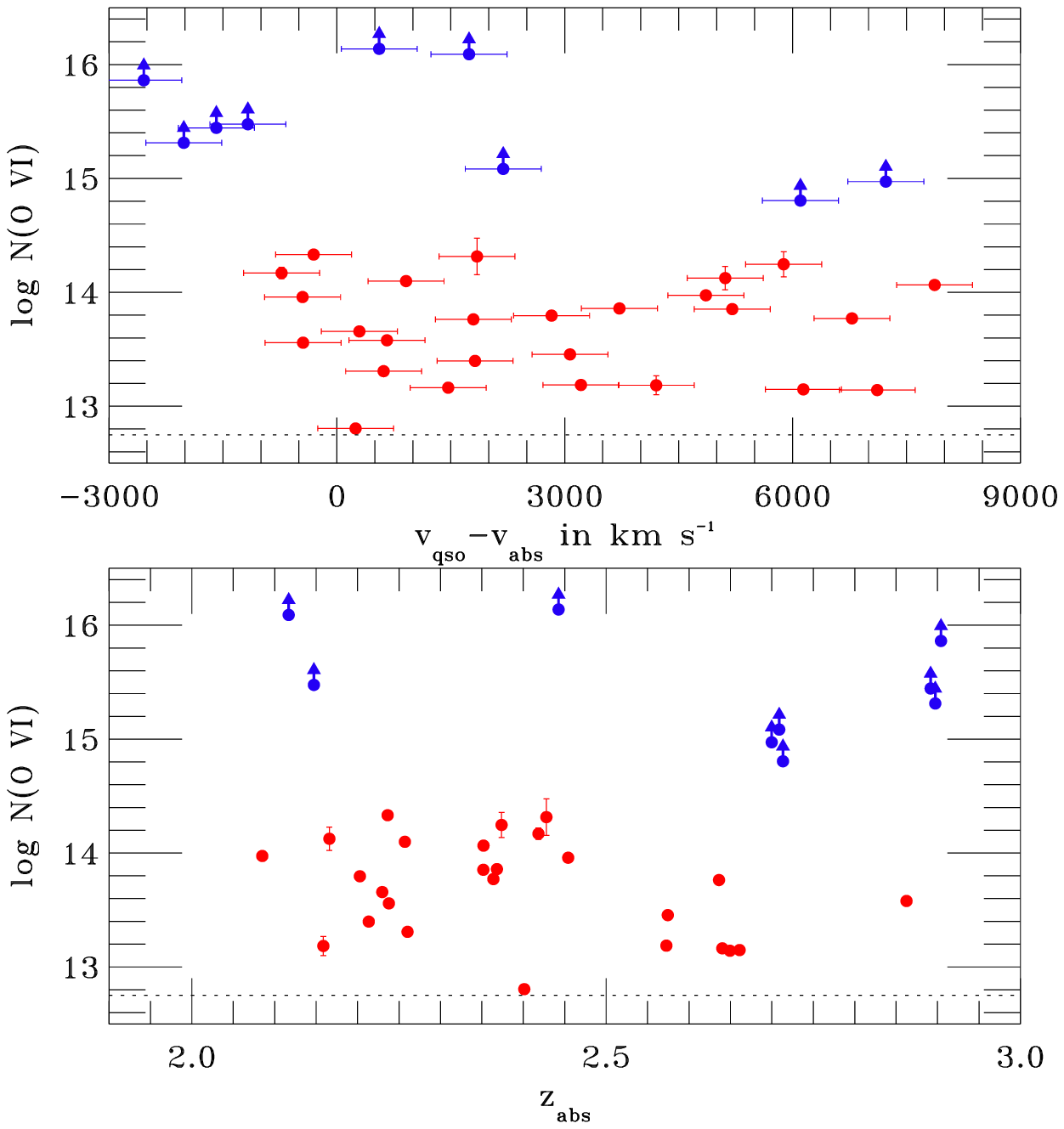}
\caption{Scatter plots comparing the {\it integrated} column density
  (over all components) of the \os\ absorbers with their proximity
  to the background quasar  $\delta v\equiv v_{\rm qso}\!-\!v_{\rm abs}$
  and with their redshift \zabs. An error of 500~\kms\ in $z_{\rm
  qso}$ is assumed. Strong absorbers are shown in blue (lower limits),
  with weak absorbers in red. 
  The distinction between the weak and strong populations is evident.
  The dashed line shows the detection limit.
  Among the weak absorbers there is no correlation between
  $N$(\os) and $\delta v$, or between $N$(\os) and $z$.} 
\end{figure}

\subsection{\os\ component column densities and $b$-values}
Proximity effects may occur at the component level,
as well as the system level. To investigate this, we compare in Figure
4 the component properties returned by {\sc VPFIT}
(column densities and line widths) of the
weak \os\ absorbers with $\delta v$. We include for comparison the
\os\ components from the intervening sample of BH05. 
As with the system column densities, there is no trend for the
component \os\ column densities to correlate with proximity to the
quasar. %* 

\begin{figure}
\includegraphics[width=9.2cm]{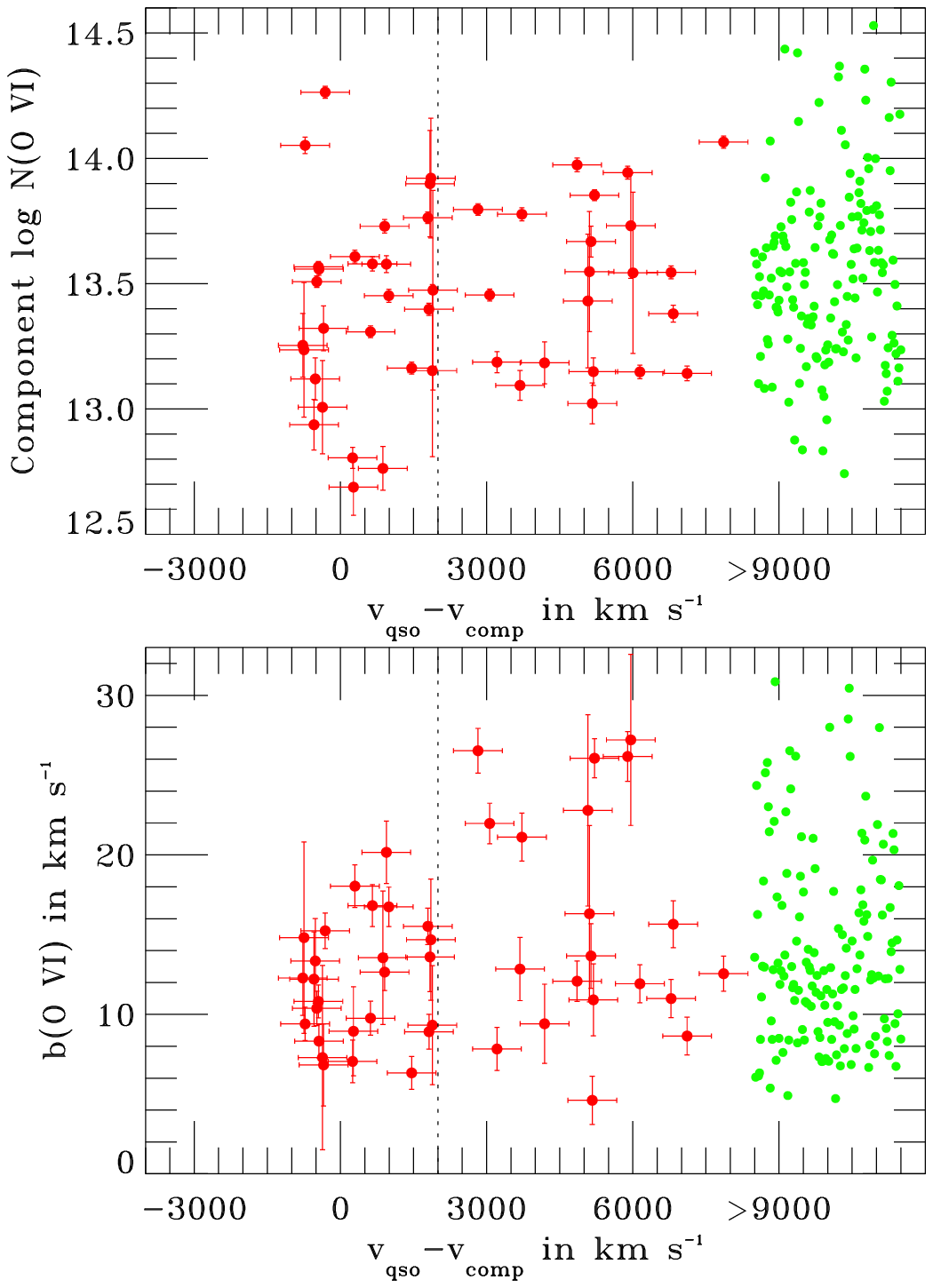}
\caption{Scatter plots comparing the column densities and line widths
  of the \os\ {\it components} with their proximity to the background
  quasar. The errors shown are returned by the {\sc VPFIT}
  profile-fitting code. Included in green is the BH05 sample of
  intervening \os\ components at $z$=2--2.5, plotted for convenience
  at $\delta v>9\,000$~\kms, with arbitrary shifts in the x-direction
  applied for clarity. Two proximate components with $b$=34$\pm$14 and
  45$\pm$18~\kms\ are not shown because of their large errors.} %*
\end{figure}

The distributions of log\,$N$ and $b$ for the \os\ components
in our sample are shown in Figure 5. We include the following
component distributions as separate curves: the components at $\delta
v\!<\!2\,000$~\kms, those at $2\,000\!<\!\delta v\!<\!8\,000$~\kms,  
and the intervening absorbers from BH05.
A two-sided K-S test finds the probability that the log\,$N$ distributions 
for the samples at $<$2\,000~\kms\ and
2\,000--8\,000~\kms\ are different (i.e. drawn from
distinct parent populations) is only 23\%. %*  
Comparing the $\delta v\!<\!2\,000$~\kms\ components to the
intervening sample, we do not find any difference valid at more than
the 1.0$\sigma$ level, %* 
with the intervening sample showing a median log\,$N$(\os) of 13.54,
and the $\delta v\!<\!2\,000$~\kms\ sample showing a median of 13.45. 
This is an important result considering the intervening sample was
measured independently in BH05, not in this analysis. %*

\begin{figure}
\includegraphics[width=9.2cm]{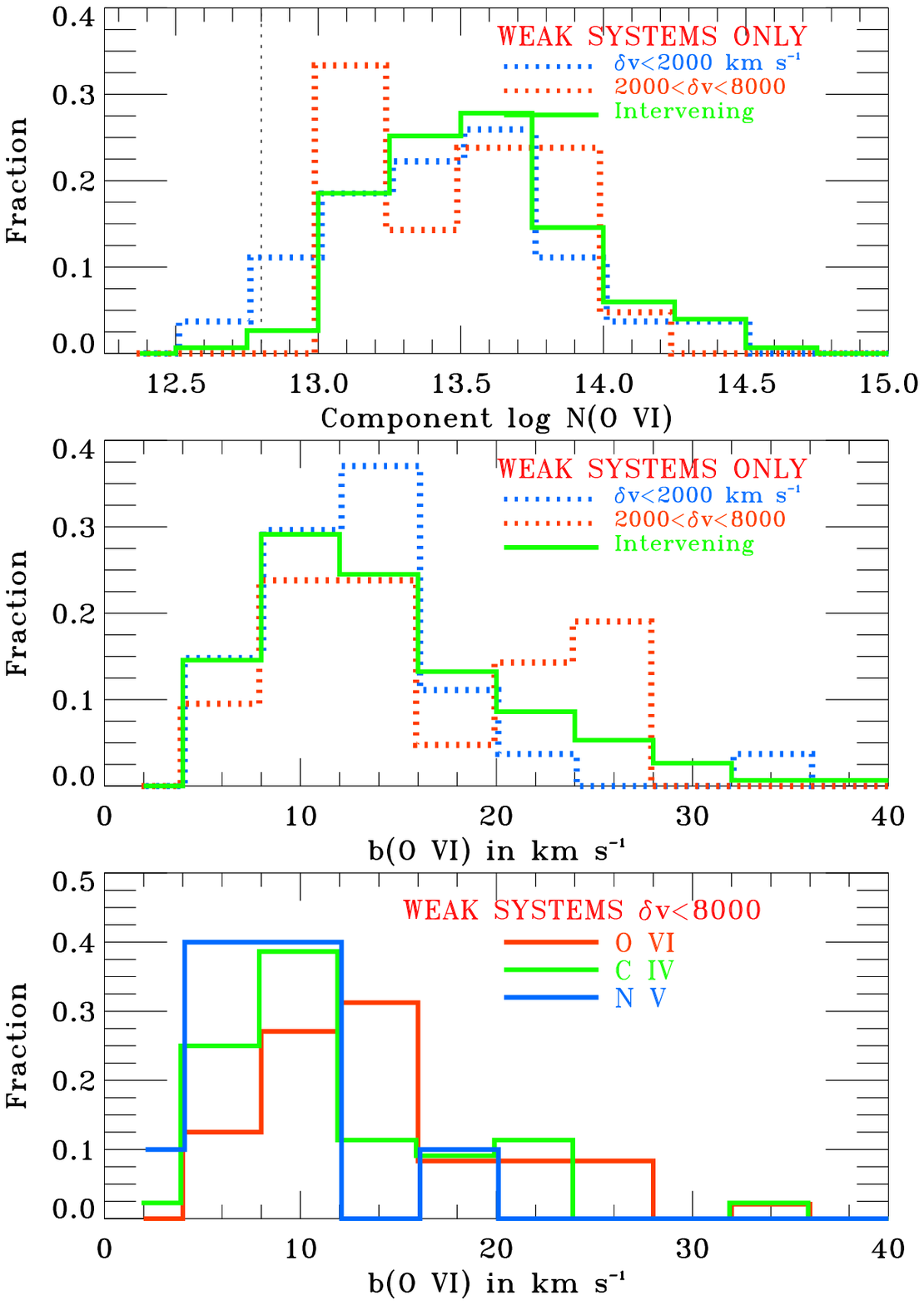}
\caption{
  Top two panels: normalized column density- and line width
  distributions of the proximate \os\ components 
  at $\delta v\!<\!2\,000$~\kms, at $2\,000\!<\!\delta v\!<\!8\,000$~\kms,
  and in the intervening sample. 
  No statistically significant differences are found, with 
  the exception of a slight excess of systems with $b$=20--30~\kms\
  in the weak sample in the 2\,000--8\,000~\kms\ bin. %*
  The bottom panel compares the line width distributions for \os, \cf,
  and \nf\ in the weak, proximate sample.}
\end{figure}

Turning to the component line widths, we find that 
among the proximate weak sample, 29 out of 48 components
(60\%) are narrow ($b\!\le\!14$~\kms)\footnote{We 
  define `narrow' as $b\!<\!14$~\kms\ since this corresponds to
  the thermal line width of an \os\ line in gas at 188\,000~K, and
  negligible \os\ is produced in gas in collisional ionization
  equilibrium (CIE) below this temperature \citep{SD93, GS07}.}. %*
Comparing the $b$-value distributions, we find that the 
probability that the population at $\delta v<2\,000$~\kms\
is different from the intervening distribution is less than 50 per cent. %*
The median $b$-value in the intervening sample is 12.7~\kms; in the
sample at $\delta v\!<\!2\,000$~\kms, it is 12.3~\kms. %*
However, an excess of systems with $b$=20--30~\kms\
is observed for components in the 2\,000--8\,000~\kms\ range, as can
be seen in the lower panel of Figure 4. %*

Finally, we compare the line width distribution of the \os\ components
in the weak proximate sample with the \cf\ and \nf\ line width
distributions (in the same sample). The line widths of \cf\ and \nf\
distribute differently than those of \os, without the strong tail
extending between 10 and 25~\kms\ seen in the \os\ distribution.
77\% of \cf\ components and 90\% of \nf\ components show %*
$b\!<\!14$~\kms. These differences indicate that it cannot be assumed
that the \cf, \nf, and \os\ arise in the same volumes of gas.

In summary, among the weak \os\ absorbers at $z$=2--3, there is no
compelling evidence for any difference in the column density and line width
distributions between the intervening and proximate \os\ populations, 
but within the proximate sample, the \os\ line widths distribute
differently than those of \nf\ and \cf. %*

\subsection{\hi\ column density}
The median and standard deviation of the logarithmic \hi\
column density in the weak proximate \os\ sample is
log\,$N$(\hi)=14.66$\pm$0.95, with a total range covered in
log\,$N$(\hi) of over four decades. The \hi/\os\ ratio 
(integrated over all components in each system)
varies enormously, taking values between 0.02 and 1500. %*  
In Figure 6 (top panel) we plot the \hi/\os\ column density ratio
against the \hi\ column density. Within both the strong and weak
samples, a trend is evident in which \hi/\os\ rises with increasing
$N$(\hi). However, the two populations are clearly distinct, in that
the weak systems show much higher \hi/\os\ ratios at a given
$N$(\hi). The clear segregation of the weak and strong systems on this plot
supports our contention that they represent two separate
populations. 

\begin{figure}
\includegraphics[width=9.2cm]{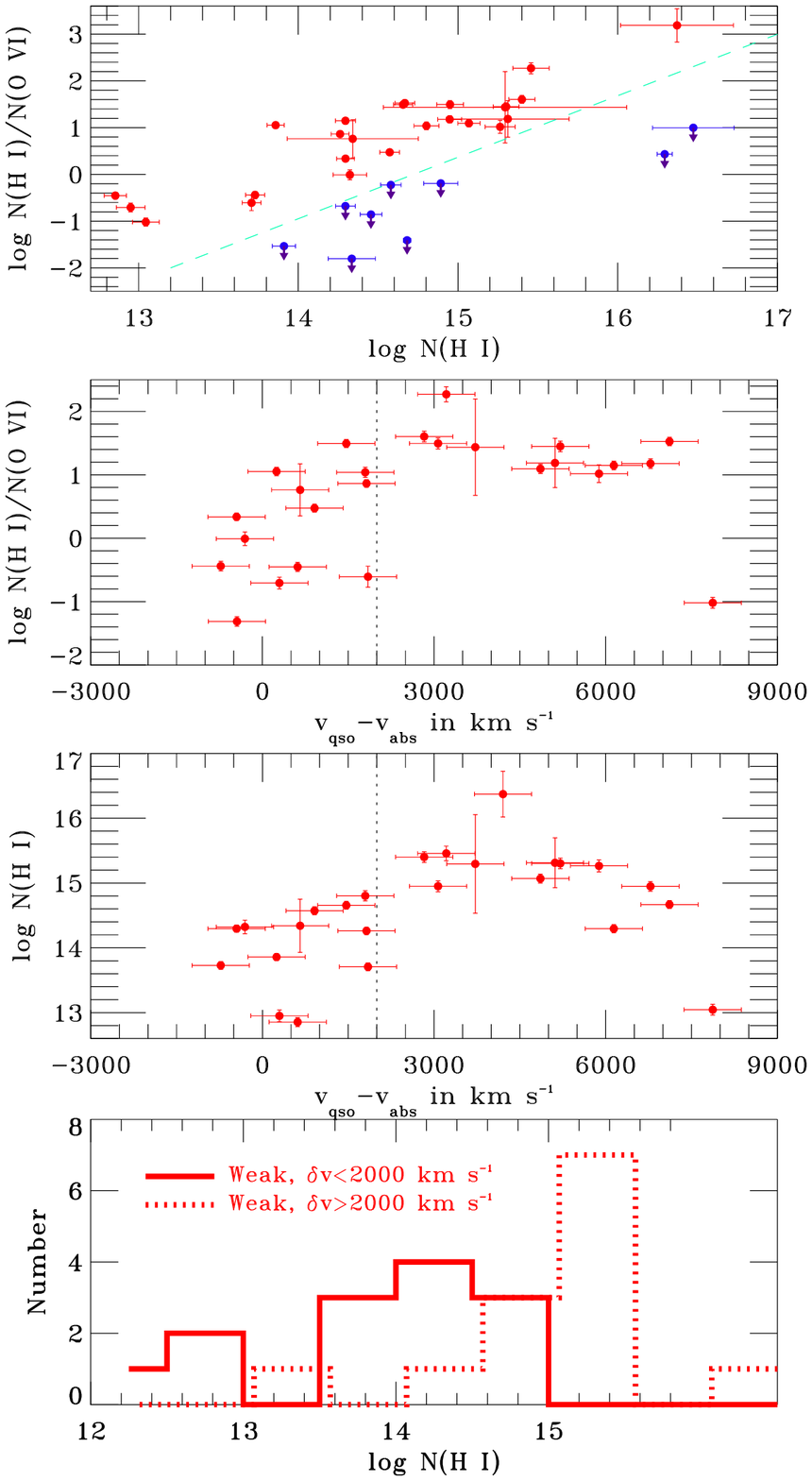}
\caption{Analysis of $N$(\hi) and the \hi/\os\ ratio in the proximate
  \os\ sample. The column densities are integrated over all components
  in each absorber. In the top panel, a dashed line delineates the locations
  of the strong (blue) and weak (red) samples in a plot of \hi/\os\ vs
  \hi. In the central two panels we show that both $N$(\hi) and \hi/\os\
  show a dependence on proximity. The bottom panel shows the histogram
  of $N$(\hi) for the weak \os\ absorbers, split into two sub-samples:
  those at $\delta v\!<\!2\,000$ and $\delta v\!>\!2\,000$~\kms. 
  A significant difference exists between these two histograms, with
  the absorbers closer in velocity to the quasar showing lower $N$(\hi).}
\end{figure}

In the lower panels of Figure 6, we explore whether the
\hi\ column density in the weak proximate sample depends on $\delta v$.
We find that $N$(\hi) distributes very differently
at $\delta v\!<\!2\,000$~\kms\ than at $\delta v\!>\!2\,000$~\kms. 
The $N$(\hi) distribution for the $\delta v\!>\!2\,000$~\kms\ sub-sample is
strongly clustered around log\,$N$(\hi)$\approx$15, whereas the distribution
for the $\delta v<2\,000$~\kms\ sub-sample is centered near
log\,$N$(\hi)$\approx$14 (see bottom panel). A K-S test shows the two
distributions differ at the 99.97\% level. %*
Consequently, there is also a difference in the mean \hi/\os\ ratio
above and below 2\,000~\kms, but this is due to the proximity effect in
$N$(\hi), not in $N$(\os), which shows no correlation with proximity
to the quasar.
The finding that $N$(\hi) correlates with $\delta v$ has an important
implication: it implies that $\delta v$ does correlate with physical
distance to the quasar. If $\delta v$ was dominated by peculiar
motions rather than Hubble flow, and so was uncorrelated to physical
proximity to the quasar, then the two $N$(\hi) distributions shown in the
bottom panel of Figure 6 would not be distinct. 
The fact that they are so different allows us to make the
general assumption that proximity in velocity does imply proximity in
distance.%* 

\subsection{\cf\ and \nf\ column densities}
The logarithmic \cf\ column densities in the proximate \os\ sample
cover over 2~dex, with a median value and standard deviation of %*
log\,$N$(\cf)=12.64$\pm$0.60. \nf\ is detected in nine
weak absorbers, with log\,$N$(\nf) between 11.54 and 13.60. %* 
%Among the weak proximate sample, we measure the median and standard
%deviation of the \cf/\os\ and \nf/\os\ ratios to be
%log\,[$N$(\cf)/$N$(\os)]=$-$0.81$\pm$0.66 and %*
%log\,[$N$(\nf)/$N$(\os)]=$-$1.81$\pm$0.70 %* 
%(where the upper limits were excluded in forming these values, and the
%column densities are integrated over all components in each system).
%These high-ion column density ratios are of use in diagnosing physical
%conditions and ionization mechanisms, although their use depends on
%whether the various ions co-exist in the same volumes of gas (see \S4).
Comparisons between the \cf\ column density and $\delta v$, and
between \nf\ column density and $\delta v$, as well as the dependence
of the high-ion column density ratios with $\delta v$, are presented in
Figure 7. In a similar fashion to \hi, $N$(\cf) in the
weak \os\ absorbers depends strongly on $\delta v$, with the absorbers
at $\delta v\!<\!2\,000$~\kms\ showing column densities clustered
around log\,$N$(\cf)$\approx$12.4, and those at $\delta v\!>\!2\,000$~\kms\
showing values centered at log\,$N$(\cf)$\approx$13.3. A two-sided K-S %*
test shows that the $N$(\cf) distributions in the two velocity intervals
differ at a significance level of 93\%. %*
This leads to a difference in the mean \cf/\os\ ratio between the two
velocity bins, with lower ratios at lower velocity.
There is no indication that either $N$(\nf) or \nf/\os\
depends on $\delta v$, though the smaller sample size of \nf\
(because of the non-detections) prevents us from making statistically
meaningful conclusions. 

\begin{figure*}
\includegraphics[width=18cm]{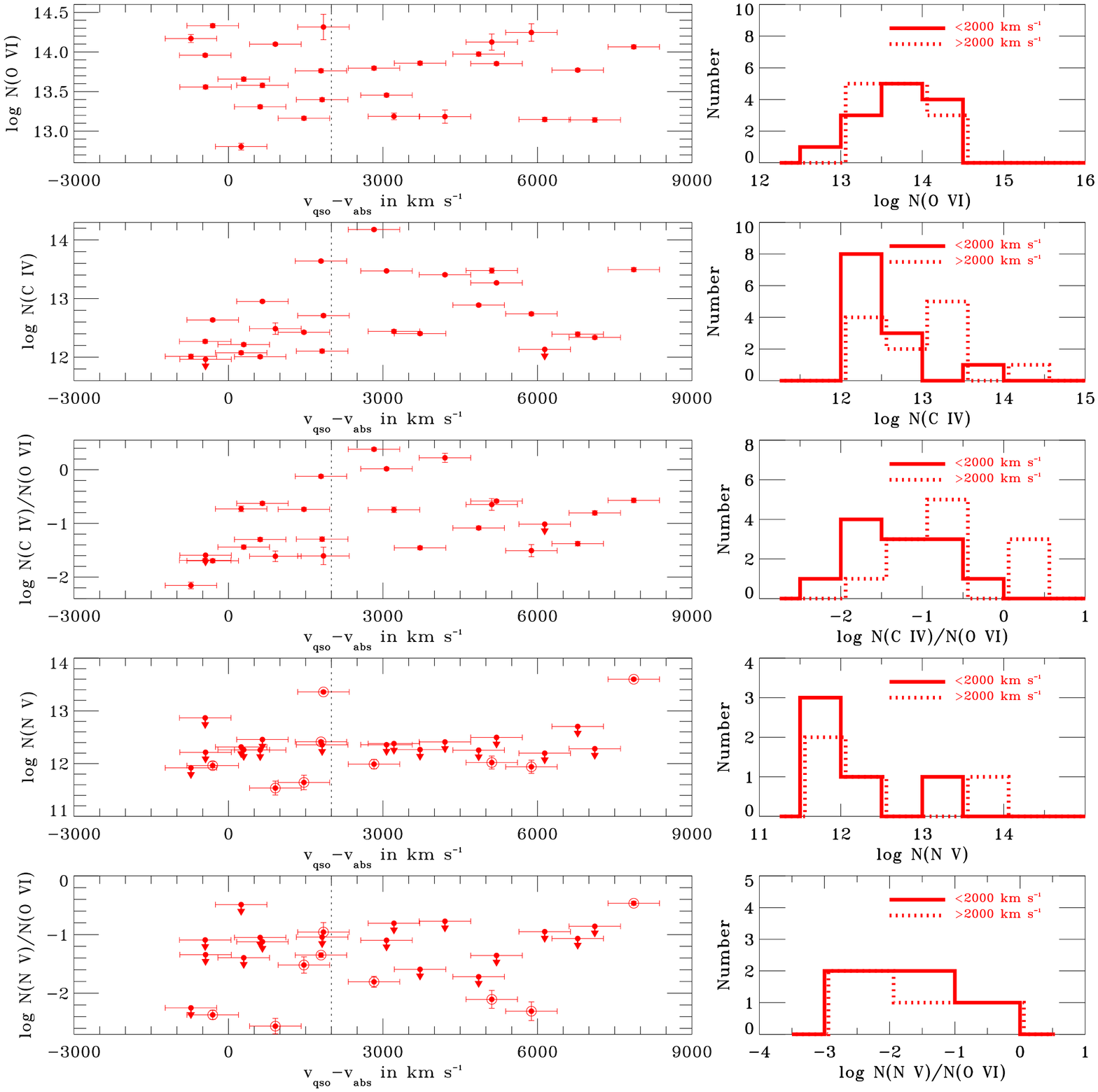}
\caption{Dependence of $N$(\os), $N$(\cf), $N$(\cf)/$N$(\os), 
  $N$(\nf), and $N$(\nf)/$N$(\os) on  $\delta v$, for the weak 
  proximate \os\ sample. Downward-pointing arrows denote
  non-detections. The right column shows the histograms of each
  quantity plotted on the left, split into sub-samples at 
  $\delta v\!<\!2\,000$ and $>$2\,000~\kms\ from the quasar. 
  In the lower two panels, the nine \nf\ detections are shown with
  open circles to distinguish them from the non-detections. Only
  detections are included in forming the histograms.
  Whereas there is no discernable effect for $N$(\os) to 
  depend on $\delta v$, $N$(\cf) distributes very differently at $<$2\,000 
  than at $>$2\,000~\kms. The \nf\ samples are too small to draw
  conclusions about whether $N$(\nf) depends on $\delta v$.}
\end{figure*}

In order to study the high-ion velocity structure more closely, we
present in Figures 8a and 8b apparent column density profiles as a
function of velocity, for each weak proximate \os\ absorber. 
The apparent column density per unit velocity in units of
ions~\sqcm~(\kms)$^{-1}$ is given by
$N_a(v)=3.768\times10^{14}\tau_a(v)/f\lambda$ \citep{SS91}, 
where $\lambda$ is the
transition rest wavelength in \AA, $f$ is the oscillator strength
\citep[taken from][]{Mo03}, and $\tau_a(v)$ is defined in \S2.2. %*. 
These figures allow the \os, \cf, \hi, and (in the nine cases where
present) \nf\ profiles to be closely compared. %*
Note that for \hi, we plot the apparent column density profile of our
best-fit model to the Lyman series absorption lines, because several
lines are used to derive our best-fit model. For the high
ions, the actual data from one of the two doublet lines are shown,
rather than the model.

In 11/24 absorbers where both \os\ and \cf\ are detected, %* 
the profiles are similar enough to suggest co-spatiality of the ions
C$^{+3}$ and O$^{+5}$. In the remaining 13 cases there are significant %*
differences in the line profiles, indicating that \os\ and \cf\ trace
different gas phases. Here a `significant difference' is either
a centroid offset of $>$10~\kms, or a case where
$b$(\cf)$<$\,$b$(\os), for which there is no single-phase solution
[since an oxygen atom is heavier than a carbon atom, $b$(\os) must be
  less than $b$(\cf) for lines arising in the same gas].
% no matter what the non-thermal broadening. %* 
In only one of the nine cases with \nf, its profile is similar in
centroid and width to the \os. However, detailed comparisons of the
\nf\ and \os\ profiles are complicated by the fact that low-optical
depth \nf\ absorption is lost in the noise, even in this high-quality
dataset, so we do not read into this result any further. 
Most significantly, offsets of order 10--30~\kms\ %*
between the centroid of the \hi\ absorption and 
the centroid of \os\ absorption are seen in 14/26 weak proximate cases. %*
These offsets are highly important, since they imply that the H$^0$
atoms do \emph{not} live in the same volume of gas as the O$^{+5}$
ions in at least half of the weak \os\ absorbers.
In other words, at least half of the weak \os\ systems are multi-phase.
Consequently, it is not possible to derive a metallicity estimate in
these multi-phase cases from the \os/\hi\ ratio, even with an
ionization correction, and single-phase photoionization models cannot
be applied. The dangers of treating multi-phase quasar
absorbers as single-phase have been pointed out before \citep{Gi94, Re01}.

\begin{figure*}
\includegraphics[width=18cm]{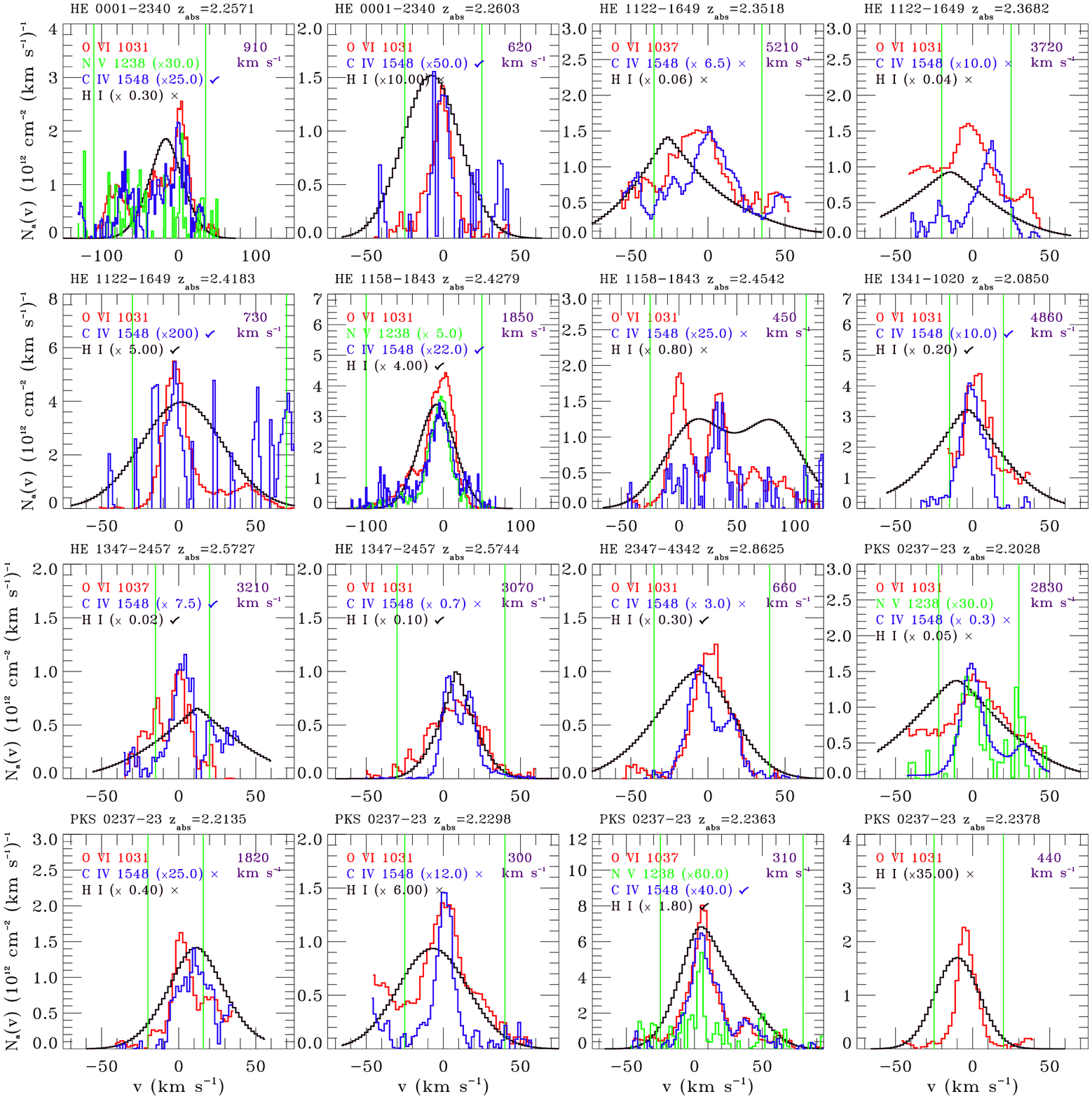}
\caption{High-ion and \hi\ apparent column density profiles for each
  weak proximate \os\ absorber. The \cf\ (blue), \hi\ (black), and
  \nf\ (green, where present) profiles have been scaled to allow their
  shape to be compared with \os. For the high ions the AOD profiles
  are formed from the data themselves. For \hi, we show the AOD
  profile of the best-fit model to the Lyman series absorption
  lines. The green vertical lines show the velocity range over which
  the AOD measurement is made. Annotated on the top-right
  of each panel is the proximity to the quasar. In cases where the
  \os\ and \cf\ profiles show similar line centers and line widths, 
  we add a tick to the \cf\ label. 
  In cases where the \os\ and \hi\ profiles show similar
  velocity centroids (but not necessarily the same width), 
  we add a tick mark to the \hi\ label.
  Cross marks next to the \cf\ and \hi\ labels indicate significant
  differences between the profiles of these ions and \os,
  e.g. centroid offsets of $>$10~\kms, indicating
  that the lines do not form in the same phase of gas as the \os.}
\end{figure*}

\addtocounter{figure}{-1}
\begin{figure*}
\includegraphics[width=18cm]{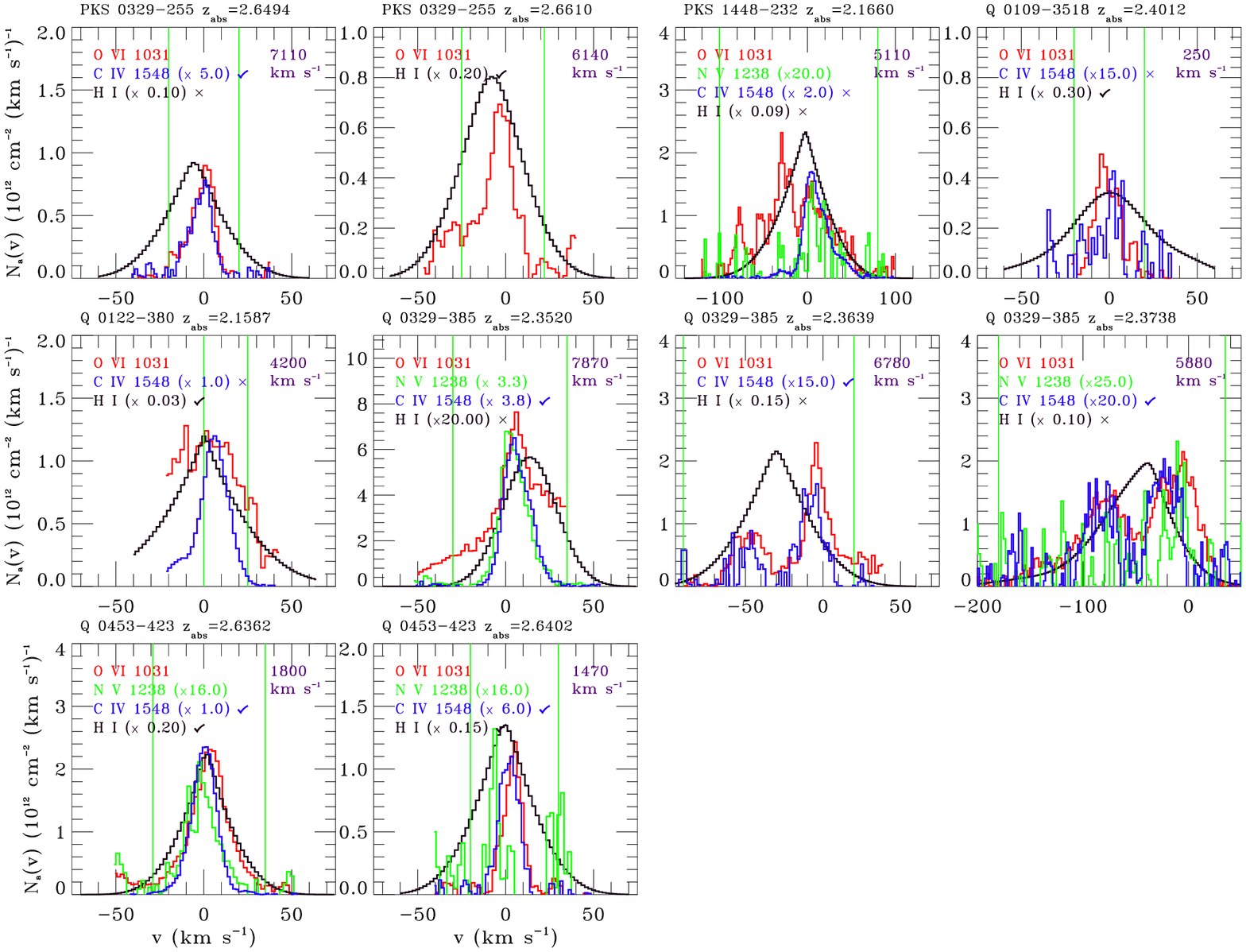}
\caption{(continued). High-ion and \hi\ apparent column density
  profiles in the weak \os\ absorbers.}
\end{figure*}

\subsection{Calculation of d$N$/d$z$ for proximate \os}
In this section we derive the incidence of proximate \os\
absorption, in terms of the traditional d$N$/d$z$,
and also in terms of the number of absorbers detected per 2\,000~\kms\
interval in $\delta v$. Note that because we are working at $z\!\gg\!0$, 
relativistic effects cause the relationship between velocity and
redshift intervals to be d$v\!\approx\!c$\,d$z$/(1+$z$).
Before calculating the incidence of proximate \os\ absorption, 
we correct the redshift path for blending with the
\lya\ forest, since \os\ is only identifiable in
unblended regions of the spectrum. To estimate the number of
\os\ absorbers that were missed due to \lya\ forest contamination, we
looked for \cf\ absorbers within 8\,000~\kms\ of \zqso\ in which
the accompanying \os\ is blended, and found five. 
These are listed in Table 5.
This method should provide a good estimate for the \os\ incompleteness,
since firstly \os\ is detected in all proximate \cf\ absorbers in which the
\os\ data are unblended, and secondly \cf\ is unblended
by the \lya\ forest, so no \cf\ systems should be missed down to a
limiting equivalent width of $\approx$2~m\AA. %*.
The limiting equivalent width for the \cf\ search is smaller
than the limiting equivalent width for the \os\ search, because \cf\
lies in a higher S/N region of the spectrum.

\begin{table}
\caption{Proximate \cf\ with blended \os}
\begin{tabular}{lccc}
\hline
QSO & \zqso\ &  \zabs\ & $\delta v$ (\kms) \\ %& Notes\\
\hline
 HE~0001-2340 &       2.2670 &       2.1870 &  7530   \\ %& ...\\
 HE~0151-4326 &       2.7890 &       2.6956 &  7580   \\ %& ...\\
 HE~1341-1020 &       2.1350 &       2.1065 &  2750   \\ %& ...\\
 HE~2347-4342 &       2.8710 &       2.8781 &  $-$550 \\ %& ...\\
 Q~0122-380   &       2.2030 &       2.1472 &  5320   \\ %& ...\\
\hline
\end{tabular}
\end{table}

The total redshift path within 8\,000~\kms\
of sixteen quasars (at $\langle z_{\rm qso}\rangle$=2.44) is 1.48. 
Because there are five `missing' \os\ systems, we estimate our
\os\ completeness to be \nsys/(\nsys+5)=84\%, %*
implying that a path equal to 0.25 is blended near \os, %*
so that the corrected redshift interval for the proximate 
\os\ search is 1.24. %*  
%Finally, including the path between $-$1\,000 and 0~\kms\
%(in which four weak systems are found) increases the total 
%unblended redshift interval to 1.39, in which \nwea\ weak systems
In this path \nwea\ weak systems (comprising 48 components) are detected. %*  
so overall d$N$/d$z$ of weak \os\ systems 
at $\delta v\!<\!8\,000$~\kms\ is 21$\pm$5, %**
where the errors on the counts are assumed to be Poissonian.
However, this is an average value, and when we break down the absorbers
into those at $\delta v\!<\!2\,000$ and those at 2\,000--8\,000~\kms, 
we find a marked change in the incidence: d$N$/d$z$ \emph{trebles}
within 2\,000~\kms, %* 
from 14$\pm$4 to 42$\pm$12\footnote{We have included the four weak \os\ %**
  absorbers at $\delta v\!<\!0$ (\zabs$>$\zqso) in the 0--2\,000~\kms\ bin, 
  since regardless of their peculiar velocity
  they must lie in front of the quasar. 
  Even without these, d$N$/d$z$ in the 0--2\,000\kms\ bin is 28$\pm$8, %*
  twice the incidence in the 2\,000--8\,000~\kms\ bin.}. %*. 
When counting components rather than systems, a similar increase in the
incidence is observed within 2\,000~\kms\ of \zqso.
These results are shown in Figure 9, and are also summarized in Table~6. 
Note how the intervening incidence of BH05 is recovered in
the range 2\,000--8\,000~\kms.
% i.e. an enhancement in d$N$/d$z$ 
%due to quasar proximity is only seen within 2\,000~\kms. 

\begin{figure}
\includegraphics[width=9.2cm]{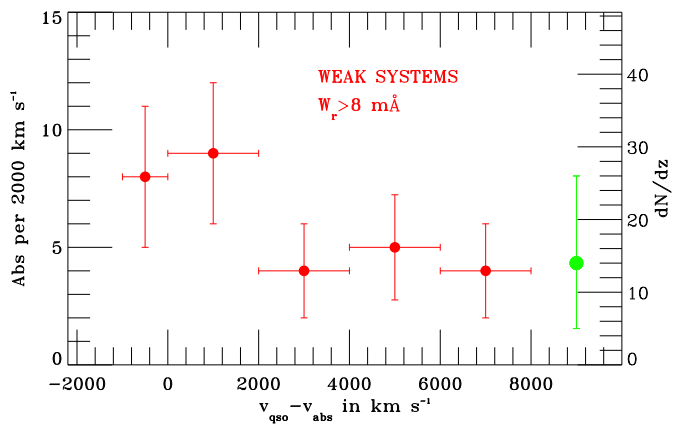}
\caption{Number of weak \os\ absorbers as a function of proximity to
the quasar. Plotted are the number of \os\ systems per 2\,000~\kms\ bin of
$\delta v$ over our sample of sixteen quasars. 
The corresponding values of d$N$/d$z$ are shown on the right,
corrected for the blended redshift path. The errors on the counts are
assumed to be Poissonian; plotted are the 1$\sigma$ errors.
We use a smaller bin for the range $-$1\,000 to
0~\kms; these absorbers lie in front of the quasar and so are included
in the 0--2\,000~\kms\ bin in our final d$N$/d$z$ calculations. %**
The incidence of intervening \os\ systems is shown as the green data
point; there is no significant difference in d$N$/d$z$ between the
absorbers at 2\,000--8\,000~\kms\ and the intervening sample.}
\end{figure}

\begin{table}
\begin{minipage}{\columnwidth}
\caption{Weak \os\ absorbers at $z$=2--3: 
Summary of d$N$/d$z$ statistics (8~m\AA\ sensitivity)}
\begin{tabular}{lccc}
\hline
Category & 
%$0\!<\!\delta v\!<\!2\,000$\footnote{Range in velocity
$|\delta v|\!<\!2\,000$\footnote{Range in velocity
  relative to \zqso\ (in \kms).} & 
$2\,000\!<\!\delta v\!<\!8\,000^{\rm a}$ 
& Intervening\footnote{Based on BH05; in the updated intervening sample
  58 systems are seen over a total $\Delta z$ of 4.09, complete down to
  log\,$N$(\os)=13.11.}\\ 
\hline
Systems    & 42$\pm$12 & 14$\pm$4 & 14$\pm$2 \\ %**
Components & 87$\pm$17 & 23$\pm$5 & 35$\pm$6 \\ %**
\hline
\end{tabular}
\end{minipage}
\end{table} %*

\subsection{Comparison of $z\approx2.5$ and $z\approx0$ absorbers}
A considerable amount of work has been invested in studying \os\ in
the low-redshift IGM using the \emph{FUSE} and \emph{HST} orbiting
observatories \citep{Be94, BT96, Tr96, Tr00, Sa02, Se04, Ri04, Pr04,
  DS05, DS08, Le06, Tr08, Co08, TC08a, TC08b}.
By comparing our results obtained
from proximate \os\ absorbers at $z$=2--3 with the $z\!\approx\!0$ results,
we can search for evolution of the highly ionized gas content of the
Universe. Since the intensity of the extragalactic UV background rises
with redshift 
\citep{HM96}, one might expect to see higher photoionized \os\ column
densities at higher redshift. Simultaneously, the fraction of baryons
in warm-hot gas is predicted to decrease with increasing redshift
\citep{Da01}, since at early cosmic epochs there has not been adequate
time for gas to accrete and virialize into dark matter-dominated
potential wells.

Based on an analysis of low-redshift ($0.15\!<\!z\!<\!0.5$) \os\ absorption
line systems and with a sensitivity limit of $W_{\rm r}\!>\!30$~m\AA, 
\citet{Tr08} report that d$N$/d$z$ is enhanced by a factor of $\approx$3 %*
at velocities $<$2\,500~\kms\ from the quasar compared to the
intervening sample, with
d$N$(\os)/d$z$=16$^{+3}_{-2}$ for the intervening systems and
d$N$(\os)/d$z$=51$^{+21}_{-15}$ for those at $\delta v<2\,500$~\kms.
Our result for weak \os\ systems at $z$=2--3 
with $W_{\rm r}\!>\!30$~m\AA\ and $\delta v\!<\!2\,000$~\kms\ is
d$N$/d$z$ is 32$\pm$10, %** 
lower than the low-redshift result by a factor of $\approx$1.6. %**
However, comparing the high- and low-redshift results for d$N$/d$z$
can be misleading, because redshift has a highly non-linear
relationship with spatial dimensions, that depends on cosmology.
With our adopted cosmology, then in the interval $z$=0.15 to $z$=0.5,
d$x$(Gpc)=3.63\,d$z$, whereas for the interval $z$=2 to $z$=3,
d$x$(Gpc)=1.22\,d$z$, %*
where d$x$ is the comoving radial distance, and where 
we have used the online cosmology calculator of \citet{Wr06}. 
Correcting for these factors to calculate the number density of weak \os\
absorbers per unit co-moving distance, a better indicator of the space
density of \os, we derive the results listed in Table 7.
Here we see evidence that the space density of \os\ absorbers 
\emph{increases} with redshift, by a factor of two between
$z$=0.15--0.5 and $z$=2--3. This increase is observed independently 
for both the proximate and intervening absorbers. %**

\begin{table}
\begin{minipage}{\columnwidth}
\caption{d$N$(\os)/d$x$ statistics (30~m\AA\ sensitivity)}
\begin{tabular}{lccc}
\hline
Category\footnote{The entries in this table list the incidence of weak \os\
  absorbers per unit comoving distance, in units of Gpc$^{-1}$;
  the entries were derived by correcting d$N$/d$z$ for cosmology, and
  are presented for both proximate and intervening
  absorbers, at low and high redshift, and for systems and components.}
 & Sample & $z$=0--0.5\footnote{Derived using results from \citet{Tr08}.} 
 & $z$=2--3 \\
 & & (Gpc$^{-1}$) & (Gpc$^{-1}$)\\
\hline
Systems     &  Proximate ($\delta v\!<\!2\,000$)
& 14$^{+6}_{-4}$ %Tr08
& 26$\pm$8\\     %**
& Intervening ($\delta v\!>\!5\,000$) 
& 4$\pm$1        %Tr08
& 8$\pm$2\footnote{Derived from BH05 result.} \\ %*
\hline
Components  & Proximate ($\delta v\!<\!2\,000$) 
& 30$^{+8}_{-6}$ %Tr08
& 71$\pm$14\\     %**
& Intervening ($\delta v\!>\!5\,000$)
& 6$\pm$1        %Tr08
& 21$\pm$3$^{\rm c}$\\ %*
\hline
\end{tabular}
\end{minipage}
\end{table} 

\section{Discussion}
The BAL, mini-BAL, and intrinsic systems, which together we classify
as strong systems based on their fully saturated \os\ absorption, strong
accompanying \os\ and \cf\ absorption, and frequent evidence for
partial coverage, are well-understood as being formed in either
QSO-driven outflows or inflows in the immediate vicinity of 
the AGN central engine \citep[see review  by][]{HF99}.  
Here we focus on the weak (narrow) proximate systems, which represent
a separate population. Is there any evidence that these weak systems
are directly photoionized by the quasar?

We report two results that %could be referred to as `\os\ proximity effects',
address this question, though their interpretation is not straightforward.
First, in our sample of proximate \os\ absorbers at $z$=2--3, there is an
enhancement by a factor of three %*
in the incidence d$N$/d$z$ of weak \os\ absorbers within 2\,000~\kms\ of the
quasar versus those in the interval 2\,000 to 8\,000~\kms.
A similar enhancement is seen near low-redshift quasars \citep{Tr08}.
While at face value this could be interpreted as a proximity effect,
in which quasars preferentially ionize nearby
clouds more often than they photoionize more distant clouds, this is
not the only explanation. The enhancement in d$N$/d$z$
could also be explained by an over-density
of galaxies near quasars, and where the \os\ absorbers are located in
the gaseous halos of these galaxies \citep{Yo82}.
It is well-known that quasars are preferentially formed in cluster
environments, particularly at high redshift \citep[][]{Cr05, Sh07}. 
This idea is supported by the observation that the internal properties
of the weak \os\ absorbers are not dependent on redshift or proximity to
the quasar.

Second, there are statistically significant differences %*
between the \hi/\os\ and \cf/\os\ ratios measured in the weak \os\
populations above and below 2\,000~\kms, with the weak
absorbers at $\delta v\!<\!2\,000$~\kms\ showing a median \hi/\os\ ratio
lower by a factor of $\approx$10, and a median \cf/\os\ ratio lower by a %*
factor of $\approx$7 than those at $2\,000\!<\!\delta v\!<\!8\,000$~\kms. %*
Again, at face value, these results could be interpreted as
proximity effects, in which gas closer to the quasar shows a higher
ionization level than intervening gas. 
However, closer inspection finds that while
$N$(\hi) and $N$(\cf) show a tendency to decrease as $\delta v$
decreases, $N$(\os) is uncorrelated with $\delta v$, i.e. 
it is the behaviour of $N$(\hi) and $N$(\cf) alone that is driving the
trends seen in the column density ratios, not the behaviour of $N$(\os).
Importantly, the \hi\ and \cf\ proximity effects imply that proximity
in velocity does correlate with proximity in distance, which supports
the idea that the velocities of the weak absorbers are dominated by
the Hubble flow.

%Because the \os\ column density distribution is independent of
%proximity, and 
Because \os-\hi-\cf\ velocity centroid offsets are
observed directly in $\approx$50 per cent of the weak systems, 
single-phase photoionization models for the \os, \cf, and \hi\ are
inadequate for at least half the absorbers in our sample. 
Indeed, the non-dependence of $N$(\os) on proximity casts doubt on
whether photoionization by the background QSO creates the \os\ at all. 
Photoionization by nearby stellar sources of radiation is
also unlikely, since such sources do not emit sufficient fluxes of
photons above 54\,eV (the {\mbox{He\,{\sc ii}}} ionization edge)
to produce the observed quantities of \os.
We cannot rule out photoionization by the quasar in every
individual case, for example the absorber at $z$=2.4183 toward
HE~1122-1649, which shows strong \os\ but barely detectable \cf, as
do several proximate \os\ absorbers reported by \citet{Go08}.
Nonetheless, $N$(\os) does not depend on proximity in the way that $N$(\hi)
and $N$(\cf) do. Thus we infer that the sizes of spheres-of-influence around
quasars, which are implied the shapes of Gunn-Petersen troughs
\citep{ZD95, Sm02}, and from studies of proximate absorption in lower
ionization species, such as \mgii\ \citep{Va08, Wi08} and \cf\
\citep*{Fo86, Ve03, Ne08}, are dependent on photon energy, and
the presence of a sphere of photons at energies above 113.9~eV (capable of
ionizing O$^{+4}$ to O$^{+5}$) has yet to be demonstrated. %*
%We ran a series of CLOUDY \citep[v96.01;][]{Fe98} photoionization
%models to several test cases where the \os\ and \hi\ line profiles are
%consistent with a single-phase model. These models assume the gas
%exists in a uniform density slab illuminated with an input QSO
%ionizing spectrum. We found the values of [O/H], [N/O], and ionization
%parameter log\,$U$ that best reproduce the observed column densities
%of \os, \cf, and \hi, assuming [O/C]=0 and taking the solar elemental
%abundances from \citet{Gr07}. 
%For the system at \zabs=2.4183 toward HE~1122-1649, we find
%[O/H]$\approx-0.5$, [N/O]$\la-0.3$, log\,$U\approx-0.2$. 
%For the system at \zabs=2.2298 toward PKS~0237-23, we find
%[O/H]$\approx-0.2$, [N/O]$\la-0.1$, log\,$U\approx-0.7$. 
%For the system at \zabs=2.2363 toward PKS~0237-23, we find
%[O/H]$\approx-1.0$, [N/O]$\la-1.0$, log\,$U\approx-0.5$. 
%These values of log\.$U$ are consistent with those found for
%intervening systems by BH05

Thus we turn to collisional ionization models.
Since the line widths of a significant fraction
($\approx$60\%) of the components in the weak \os\ absorbers %*
are low enough ($b\!<\!14$~\kms) to imply gas temperatures below 
188\,000~K, collisional ionization {\it equilibrium} can be ruled out,
because essentially no \os\ is produced in gas in CIE at these
temperatures \citep{SD93, GS07}. 
Indeed, the narrow line widths of many intergalactic \os\
absorbers at $z\!\approx\!2$ have led various authors to conclude that
photoionization is the origin mechanism \citep{Ca02, Be02, Lv03, Be05,
Re06, Lo07}.
However, \emph{non-equilibrium} collisional ionization models cannot
be ruled out so easily. Indeed one expects that collisionally ionized gas
at `coronal' temperatures of a few $\times10^5$~K, where \os\
is formed through collisions, will be in a non-equilibrium state.
This is because the peak of the interstellar cooling curve exists at
these temperatures, and so the cooling
timescales are short. When the cooling times are shorter than the
recombination timescales, `frozen-in' ionization can result at
temperatures well below those at which the ions exist in equilibrium 
\citep{Ka73, SM76, EC86}, provided that there is a source of
$\sim10^6$~K gas in the first place.

There are at least two physical reasons why collisionally-ionized,
million-degree regions of interstellar and intergalactic gas
could arise in the high-$z$ Universe. 
The first is the (hot-mode) accretion and shock heating of gas falling into
potential wells \citep{BD03, Ke05, DB06}, 
a process which is incorporated into cosmological
hydrodynamical simulations \citep{CO99, Da01, FB03, Ka05, CF06}, 
and creates what is referred to as the Warm-Hot Intergalactic
Medium. 
However, these models generically predict that the fraction of
all baryons that exist in the WHIM rises from essentially zero at
$z$=3 to 30--50\% at $z$=0, so \emph{little WHIM is expected at the
redshifts under study here}.
The second reason is the presence of galactic-scale outflows, which
due to the energy input from supernovae are likely to contain (or even
be dominated by) hot, highly ionized gas 
\citep[see recent models by][]{OD06, Fa07, KR07, Sa08}.
There is strong observational evidence for outflows at redshifts of
$\approx$2--3 \citep{He02}, including blueshifted absorption 
in the spectra of Lyman break galaxies \citep{Pe00, Pe02, Sh03}, 
the presence of metals in the low-density,
photoionized IGM \citep[the \lya\ forest; e.g.][]{Ar04, Ag05, Ag08},
though see \citet{Sy07}, and the presence of super-escape velocity
\cf\ components in the spectra of damped \lya\ (DLA) galaxies \citep{Fo07b}.
In addition, collisionally ionized gas in
galactic halos at high redshift has been seen directly through
detections of \os\ and \nf\ components in DLAs \citep{Fo07a},
albeit with much broader system velocity widths than in the proximate
absorbers discussed here. \os\ absorbers with lower \hi\ column
density may probe the outer reaches of such halos or `feedback zones'
\citep[BH05;][]{Si02}.

We explore the ability of non-equilibrium collisional ionization
models to reproduce the data in our proximate \os\ sample in 
Figure 10, which shows the \cf/\os\ vs
\nf/\os\ ratio-ratio plane. We take the isobaric non-equilibrium model
at log\,$T$=5.00 (consistent with the observed \os\ component line
widths) from \citet{GS07}, computed using solar abundances, and then find the
values of [N/O] and [C/H] that are required to reproduce the
observations. The models for 0.1 and 0.01 solar absolute abundances,
and for the isochoric case, predict similar values.
The non-equilibrium models are not capable of reproducing the observed
ratios when solar relative elemental abundances are used. 
However, if [N/O] takes values between $-$1.8 and 0.4, %*
and [C/O] between $-$1.9 and 0.6, it is possible to explain 
the observed column densities in each weak proximate system
with a non-equilibrium collisional ionization model.
Note that we have not corrected these model predictions for the effect
of photoionization by the extragalactic background; the model
represents the pure collisional ionization case.
Hybrid collisional+photo-ionization models would help in making
progress in this area \citep[see discussion in][]{Tr08}.

\begin{figure}
\includegraphics[width=9.2cm]{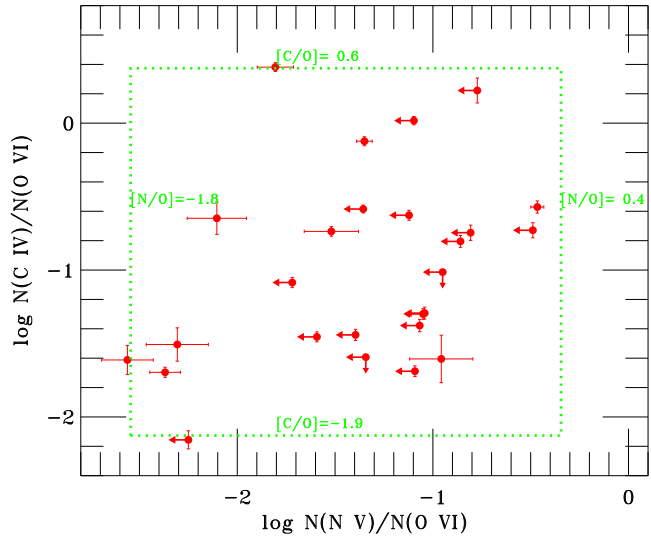}
\caption{Comparison of predictions of non-equilibrium collisional
  ionization models with the observed column density ratios in our weak
  proximate sample. Non-detections of \nf\ (\cf) lead to upper limits
  on the x- (y-) axis. We take the predictions from the isobaric
  non-equilibrium collisional ionization model of \citet{GS07} for an
  absorber at $T$=$10^5$~K (higher temperatures are ruled out by the component
  line widths).  The model is fairly insensitive to changes in the
  overall abundance level, but is sensitive to non-solar {\it
  relative} abundances. The dashed green lines show the range of
  values of [N/O] and [C/O] that are required to fit the proximate
  data with this collisional ionization model.}
\end{figure}

\section{Summary}
We have presented a study of proximate ($\delta v<8\,000$~\kms)
\os\ absorbers in the spectra of sixteen quasars observed at high
signal-to-noise and 6.6~\kms\ resolution with VLT/UVES. 
The quasars are at redshifts between 2.14 and 2.87. %*
We found \nsys\ proximate \os\ absorbers
comprising \ncomp\ components. We used component fitting
software to determine the properties of the absorption,
studied the statistical properties of the sample, and
addressed the ionization conditions in the gas. Our study has
produced the following results.

\begin{enumerate}
\item \Nstr\ of the \nsys\ \os\ systems are strong, with log\,$N(\os)\ga15.0$. 
  These absorbers all show detections of \cf\ and \nf,
  and show either broad, fully saturated, mini-BAL-type
  \os\ absorption troughs extending over tens to hundreds of \kms,
  or evidence for partial coverage in the high-ion profiles. 
  A mean of 6.0 \os\ components are seen in the strong
  absorbers. The strong absorbers are formed either in QSO-driven
  outflows or in the immediate vicinity of the quasar.

\item \nwea\ of the \nsys\ \os\ systems are weak, and show
  log\,$N(\os)\!<\!14.5$. These systems are narrow, with a median
  velocity width $\Delta v_{90}$ of 42~\kms. %*
  The weak systems all contain detectable \hi, all but two show \cf,
  and 9/\nwea\ show \nf. We find a mean of 1.9 \os\ components per %*
  weak system. Approximately 60 per cent of %*
  these \os\ components have Doppler $b$-values $<$14~\kms. 
  The gas in these narrow components is constrained by the line widths
  to be at $T\!<\!188\,000$~K.

\item Among the weak sample there is no correlation between the \os\
  column density and the proximity to the quasar. This is true when
  using either the component-level or the system-level column
  densities. Two-sided K-S tests confirm that the column density
  distributions of the intervening and weak proximate \os\ samples are
  statistically indistinguishable. %*
  The \os\ column density also does not correlate with redshift.

\item  There is a difference between the line width distributions
  of the \os\ components versus the distributions of \cf\ and \nf, in
  the weak proximate sample. The \cf\ and \nf\ distributions are
  strongly peaked below $b$=10~\kms\ and do not show the tail of
  absorbers with $b$-values between 10 and 25~\kms\ seen in the
  \os\ distribution. %*

\item There is a significantly lower $N$(\hi) in the weak \os\ 
  absorbers with $\delta v\!<\!2\,000$~\kms\ than in those with 
  $\delta v\!>\!2\,000$~\kms. 
  The difference in the median $N$(\hi) between these two samples is
  $\sim$1.0~dex.  We interpret this as a manifestation of
  the \hi\ proximity effect, and it leads to a trend in which the \hi/\os\
  ratio is lower for absorbers within 2\,000~\kms\ of the QSO than for
  those at higher $\delta v$. An identical trend is observed for \cf,
  with the median $N$(\cf) lower by $\approx$0.8~dex at $\delta
  v\!<\!2\,000$~\kms\ than at $\delta v\!>\!2\,000$~\kms. %*
  Thus \cf\ and \hi\ behave differently than \os.

\item Apparent column density profile comparisons show that \os\ and
  \cf\ show significantly different velocity structure in 13 of the 24
  weak \os\ absorbers where \cf\ is detected. %*
  Offsets between the centroid of the \hi\ model and the \os\
  components are seen in 14/26 weak proximate cases. %*
  \emph{Therefore at least half of the weak \os\ systems are multi-phase.}

\item Down to a limiting rest equivalent width of 8~m\AA, the number
  density of weak \os\ systems within 2\,000~\kms\ of the background
  quasar is d$N$/d$z$=42$\pm$12 (87$\pm$17 for components). %**
  Between 2\,000 and 8\,000~\kms\ in $\delta v$, d$N$/d$z$ falls to
  14$\pm$4 (23$\pm$5 for components), equal to the incidence of
  intervening \os\ measured by BH05. %**
  Thus an enhancement in d$N$/d$z$ by a factor of three %*
  is seen at $\delta v\!<\!2\,000$~\kms.

\item We compare the incidence d$N$/d$z$ of \os\ absorbers at
  high ($z$=2--3) and low ($z$=0--0.5) redshift.
  After correcting for cosmology, the incidence of 
  \os\ systems with $W_{\rm r}\!>\!30$~m\AA\ \emph{per unit comoving
  distance} at $z$=2--3 is twice the value at $z$=0--0.5. %**
  This increase is seen for both proximate and intervening absorbers.
 
\item 
  %$Because the \os\ column densities in the weak \os\ sample at
  %$z$=2--3 show no dependence on proximity to the quasar, and
  Because \os-\cf-\hi\ velocity offsets are observed in half the sample,
  single-phase photoionization models are (at least for these cases) 
  unable to explain the data. Indeed, the non-dependence of $N$(\os) on
  proximity casts doubt on whether photoionization creates the \os\ at
  all. Instead, we propose that the weak \os\ absorbers
  trace collisionally-ionized regions that exist as part of
  multi-phase galactic or protogalactic structures. 
  In order to reconcile the collisional
  ionization hypothesis with the narrow line widths measured in
  approximately 60 per cent of the \os\ components, the gas must be
  out of CIE. This can be explained by a scenario in which initially
  hot gas is cooling down through the coronal regime at $T\!\sim\!10^5$~K. 
  The observed high-ion column density ratios and line widths in the
  proximate \os\ absorbers can be explained by models of non-equilibrium
  collisionally ionized gas at $T\!\approx\!10^5$~K, but only if the
  relative elemental abundances are non-solar: we require 
  [N/O] to be in the range $-$1.8 to 0.4 and
  [C/O] between $-$1.9 and 0.6, to explain the data with these models. %* 

\item Future \os\ surveys do not need to remove absorbers (and thus
  compromise the sample size) that happen to lie at
  2\,000--8\,000~\kms\ relative to the quasar. A better approach would
  be to remove those showing either $N$(\os)$\ga$15.0 or evidence for
  partial coverage.
 
\end{enumerate}

\section*{Acknowledgments}
AJF gratefully acknowledges the support of a Marie Curie
Intra-European Fellowship (contract MEIF-CT-2005-023720) 
awarded under the European Union's Sixth
Framework Programme, which partially funded his contribution to this paper.
We thank Bastien Aracil for his continuum fits to the UVES data, and
Bob Carswell for making his {\sc VPFIT} software publically available.
We thank the anonymous referee for useful comments.
% and Todd Tripp for comments on the manuscript.

% TABLES
\clearpage
\setcounter{table}{0}
\begin{table*}
\centering
\begin{minipage}{140mm}
\caption{Proximate \os\ absorbers in the UVES Large Programme}
\begin{tabular}{lcccc cccc}
\hline
QSO & $z_{\rm qso}$ & $z_{\rm abs}$ & $\delta v$\footnote{Velocity separation between absorber and quasar, in~\kms. A negative value indicates the absorber is at higher redshift than the quasar. Errors are $\sim$500~\kms, reflecting the uncertainties in $z_{\rm qso}$.} & $\Delta v_{90}(\os)$\footnote{\os\ line width containing the central 90 per cent of the integrated optical depth, in \kms. Only given for the weak absorbers, since it cannot be measured for a saturated line.} & log\,$N$(\os)\footnote{The column densities in this table (with $N$ in cm$^{-2}$) were determined by summing the column densities of the individual components in our Voigt profile fits, with the component errors added in quadrature. For saturated absorbers, we present lower limits on the column density. The symbol $\ddag$ denotes absorbers with evidence for partial coverage of the continuum source; these column densities should be considered lower limits. For non-detections of \cf\ or \nf, we present a 3$\sigma$ upper limit to log\,$N$, based on the null measurement of the equivalent width. Absorbers marked $\dag$ are not formally detected at 3$\sigma$ significance, but a fit was executed nonetheless because of weak features in the line profiles at exactly the same redshift as \os\ (see Fig 1).} & log\,$N$(\nf) & log\,$N$(\cf) & log\,$N$(\hi)\\
\hline
{\bf Strong}\\
   HE 0151-4326 &   2.789 & 2.6998 &    7230 &                  ... &          14.97$\pm$0.04$\ddag$ &          14.13$\pm$0.01$\ddag$ &          13.33$\pm$0.10$\ddag$ &                 14.29$\pm$0.05 \\
   HE 0151-4326 &   2.789 & 2.7134 &    6100 &                  ... &          14.81$\pm$0.01$\ddag$ &          13.95$\pm$0.05$\ddag$ &                 13.39$\pm$0.01 &                 14.58$\pm$0.08 \\
   HE 1158-1843 &   2.449 & 2.4426 &    560 &                  ... &                       $>$16.14 &          14.76$\pm$0.50$\ddag$ &          14.44$\pm$0.01$\ddag$ &                 14.33$\pm$0.20 \\
   HE 1341-1020 &   2.135 & 2.1169 &    1740 &                  ... &                       $>$16.09 &          14.78$\pm$0.16$\ddag$ &          14.26$\pm$0.03$\ddag$ &                 14.68$\pm$0.03 \\
   HE 1341-1020 &   2.135 & 2.1473 & $-$1170 &                  ... &                       $>$15.48 &                       $>$14.80 &                       $>$14.81 &                 16.47$\pm$0.34 \\
   HE 2347-4342 &   2.871 & 2.8916 & $-$1590 &                  ... &                       $>$15.44 &          14.20$\pm$0.01$\ddag$ &                 13.94$\pm$0.01 &                 13.91$\pm$0.08 \\
   HE 2347-4342 &   2.871 & 2.8972 & $-$2020 &                  ... &                       $>$15.31 &                       $>$14.66 &                       $>$14.72 &                 14.45$\pm$0.07 \\
   HE 2347-4342 &   2.871 & 2.9041 & $-$2540 &                  ... &                       $>$15.86 &                 13.72$\pm$0.01 &                       $>$14.69 &                 16.29$\pm$0.05 \\
   PKS 0329-255 &   2.736 & 2.7089 &    2190 &                  ... &                       $>$15.08 &                 13.28$\pm$0.01 &                 13.60$\pm$0.01 &                 14.89$\pm$0.14 \\
{\bf Weak}\\
   HE 0001-2340 &   2.267 & 2.2571 &    910 &            109$\pm$1 &                 14.10$\pm$0.01 &           11.54$\pm$0.13$\dag$ &                 12.49$\pm$0.10 &                 14.57$\pm$0.05 \\
   HE 0001-2340 &   2.267 & 2.2603 &    620 &             28$\pm$1 &                 13.31$\pm$0.02 &                       $<$12.26 &           12.01$\pm$0.03$\dag$ &                 12.85$\pm$0.07 \\
   HE 1122-1649 &   2.410 & 2.3518 &    5210 &             60$\pm$1 &                 13.85$\pm$0.01 &                       $<$12.50 &                 13.27$\pm$0.01 &                 15.30$\pm$0.11 \\
   HE 1122-1649 &   2.410 & 2.3682 &    3720 &             38$\pm$1 &                 13.86$\pm$0.02 &                       $<$12.27 &                 12.40$\pm$0.02 &                 15.29$\pm$0.91 \\
   HE 1122-1649 &   2.410 & 2.4183 & $-$730 &             61$\pm$1 &                 14.17$\pm$0.05 &                       $<$11.92 &           12.01$\pm$0.04$\dag$ &                 13.73$\pm$0.05 \\
   HE 1158-1843 &   2.449 & 2.4279 &    1850 &             78$\pm$1 &                 14.32$\pm$0.16 &                 13.36$\pm$0.02 &                 12.71$\pm$0.02 &                 13.71$\pm$0.05 \\
   HE 1158-1843 &   2.449 & 2.4542 & $-$450 &             98$\pm$1 &                 13.96$\pm$0.02 &                       $<$12.87 &           12.27$\pm$0.03$\dag$ &                 14.30$\pm$0.06 \\
   HE 1341-1020 &   2.135 & 2.0850 &    4860 &             30$\pm$1 &                 13.97$\pm$0.02 &                       $<$12.26 &                 12.89$\pm$0.01 &                 15.07$\pm$0.08 \\
   HE 1347-2457 &   2.611 & 2.5727 &    3210 &             26$\pm$1 &                 13.19$\pm$0.04 &                       $<$12.38 &                 12.44$\pm$0.03 &                 15.46$\pm$0.15 \\
   HE 1347-2457 &   2.611 & 2.5744 &    3070 &             50$\pm$1 &                 13.45$\pm$0.02 &                       $<$12.36 &                 13.47$\pm$0.01 &                 14.95$\pm$0.10 \\
   HE 2347-4342 &   2.871 & 2.8625 &    660 &             36$\pm$2 &                 13.58$\pm$0.02 &                       $<$12.46 &                 12.95$\pm$0.01 &                 14.34$\pm$0.52 \\
    PKS 0237-23 &   2.233 & 2.2028 &    2830 &             47$\pm$1 &                 13.80$\pm$0.02 &           11.99$\pm$0.09$\dag$ &                 14.18$\pm$0.01 &                 15.40$\pm$0.10 \\
    PKS 0237-23 &   2.233 & 2.2135 &    1820 &             29$\pm$1 &                 13.40$\pm$0.02 &                       $<$12.36 &                 12.10$\pm$0.03 &                 14.26$\pm$0.04 \\
    PKS 0237-23 &   2.233 & 2.2298 &    300 &             51$\pm$1 &                 13.66$\pm$0.02 &                       $<$12.26 &                 12.22$\pm$0.02 &                 12.95$\pm$0.11 \\
    PKS 0237-23 &   2.233 & 2.2363 & $-$310 &             64$\pm$1 &                 14.33$\pm$0.02 &           11.96$\pm$0.07$\dag$ &                 12.64$\pm$0.02 &                 14.32$\pm$0.14 \\
    PKS 0237-23 &   2.233 & 2.2378 & $-$440 &             26$\pm$1 &                 13.56$\pm$0.02 &                       $<$12.22 &                       $<$11.97 &                 12.24$\pm$0.08 \\
   PKS 0329-255 &   2.736 & 2.6494 &    7110 &             24$\pm$1 &                 13.14$\pm$0.02 &                       $<$12.28 &                 12.34$\pm$0.02 &                 14.67$\pm$0.05 \\
   PKS 0329-255 &   2.736 & 2.6610 &    6140 &             36$\pm$1 &                 13.15$\pm$0.02 &                       $<$12.20 &                       $<$12.13 &                 14.30$\pm$0.06 \\
   PKS 1448-232 &   2.220 & 2.1660 &    5110 &            121$\pm$1 &                 14.13$\pm$0.10 &           12.02$\pm$0.12$\dag$ &                 13.48$\pm$0.04 &                 15.31$\pm$0.50 \\
    Q 0109-3518 &   2.404 & 2.4012 &    250 &             27$\pm$2 &                 12.81$\pm$0.04 &                       $<$12.32 &           12.08$\pm$0.03$\dag$ &                 13.86$\pm$0.05 \\
     Q 0122-380 &   2.203 & 2.1587 &    4200 &             22$\pm$1 &                 13.18$\pm$0.08 &                       $<$12.41 &                 13.41$\pm$0.01 &                 16.37$\pm$0.46 \\
     Q 0329-385 &   2.440 & 2.3520 &    7870 &             36$\pm$1 &                 14.06$\pm$0.02 &                 13.60$\pm$0.02 &                 13.49$\pm$0.03 &                 13.05$\pm$0.10 \\
     Q 0329-385 &   2.440 & 2.3639 &    6780 &             60$\pm$3 &                 13.77$\pm$0.02 &                       $<$12.70 &                 12.39$\pm$0.03 &                 14.95$\pm$0.08 \\
     Q 0329-385 &   2.440 & 2.3738 &    5880 &            148$\pm$1 &                 14.25$\pm$0.11 &           11.94$\pm$0.12$\dag$ &                 12.74$\pm$0.03 &                 15.26$\pm$0.12 \\
     Q 0453-423 &   2.658 & 2.6362 &    1800 &             42$\pm$1 &                 13.76$\pm$0.02 &                 12.41$\pm$0.03 &                 13.64$\pm$0.01 &                 14.80$\pm$0.09 \\
     Q 0453-423 &   2.658 & 2.6402 &    1470 &             23$\pm$1 &                 13.16$\pm$0.02 &           11.64$\pm$0.14$\dag$ &                 12.43$\pm$0.01 &                 14.66$\pm$0.05 \\
\hline
\end{tabular}
\end{minipage}
\end{table*}

\begin{table}
\centering
\scriptsize
\begin{minipage}{80mm}
\caption{Component fit results to weak \os\ absorbers}
\begin{tabular}{lcccc c}
\hline
QSO & $z_{\rm abs}$ & Ion & $v_0$ (\kms) & $b$ (\kms)
& log\,($N$ in \sqcm) \\
\hline
HE 0001-2340 &  2.2571 &    \os\ & $-$81.0$\pm$1.1 & 16.7$\pm$1.2 & 13.45$\pm$0.02 \\
         ... &     ... &     ... & $-$31.7$\pm$1.4 & 20.2$\pm$2.0 & 13.58$\pm$0.03 \\
         ... &     ... &     ... &    4.0$\pm$1.1 & 12.6$\pm$1.2 & 13.73$\pm$0.02 \\
         ... &     ... &     ... &    39.4$\pm$2.2 & 13.6$\pm$4.2 & 12.76$\pm$0.09 \\
         ... &     ... &    \cf\ & $-$72.8$\pm$2.1 & 12.5$\pm$3.2 & 11.93$\pm$0.09 \\
         ... &     ... &     ... & $-$28.6$\pm$6.6 & 21.5$\pm$12 & 11.99$\pm$0.21 \\
         ... &     ... &     ... &    0.3$\pm$1.5 &  9.6$\pm$2.1 & 12.09$\pm$0.12 \\
         ... &     ... &    \nf\ &    3.3$\pm$1.9 &  4.0$\pm$4.0 & 11.54$\pm$0.13 \\
         ... &     ... &    \hi\ & $-$16.7$\pm$1.1 & 32.8$\pm$1.1 & 14.57$\pm$0.01 \\
         ... &  2.2603 &    \os\ & $-$0.9$\pm$1.0 &  9.8$\pm$1.1 & 13.31$\pm$0.02 \\
         ... &     ... &    \cf\ & $-$2.5$\pm$1.4 &  4.0$\pm$2.3 & 11.42$\pm$0.10 \\
         ... &     ... &    \hi\ & $-$7.7$\pm$1.2 & 25.3$\pm$1.4 & 12.85$\pm$0.02 \\
HE 1122-1649 &  2.3518 &    \os\ & $-$7.5$\pm$1.1 & 26.1$\pm$1.2 & 13.85$\pm$0.01 \\
         ... &     ... &    \cf\ & $-$50.9$\pm$1.3 & 17.6$\pm$1.5 & 12.52$\pm$0.03 \\
         ... &     ... &     ... & $-$2.3$\pm$1.1 & 21.7$\pm$1.4 & 12.90$\pm$0.02 \\
         ... &     ... &     ... &    58.6$\pm$1.8 & 34.4$\pm$2.9 & 12.72$\pm$0.03 \\
         ... &     ... &     ... &    117.5$\pm$1.1 &  7.4$\pm$1.2 & 12.30$\pm$0.03 \\
         ... &     ... &    \hi\ & $-$30.8$\pm$1.5 & 33.3$\pm$1.4 & 15.06$\pm$0.03 \\
         ... &     ... &     ... & $-$1.6$\pm$2.0 & 54.4$\pm$1.5 & 14.89$\pm$0.04 \\
         ... &     ... &     ... &    113.1$\pm$1.2 & 27.8$\pm$1.2 & 13.86$\pm$0.02 \\
         ... &  2.3682 &    \os\ & $-$2.8$\pm$1.1 & 21.1$\pm$1.5 & 13.78$\pm$0.02 \\
         ... &     ... &     ... &    35.4$\pm$1.5 & 12.8$\pm$2.0 & 13.09$\pm$0.06 \\
         ... &     ... &    \cf\ &     9.2$\pm$1.1 & 12.2$\pm$1.1 & 12.40$\pm$0.02 \\
         ... &     ... &    \hi\ & $-$47.4$\pm$64.5 & 53.4$\pm$18 & 14.45$\pm$1.07 \\
         ... &     ... &     ... & $-$12.3$\pm$4.0 & 43.3$\pm$2.8 & 15.23$\pm$0.18 \\
         ... &  2.4183 &    \os\ & $-$3.1$\pm$1.1 &  9.4$\pm$1.0 & 14.05$\pm$0.03 \\
         ... &     ... &     ... &    19.2$\pm$2.8 & 14.8$\pm$ 9 & 13.24$\pm$0.27 \\
         ... &     ... &     ... &    44.9$\pm$2.7 & 12.3$\pm$2.4 & 13.25$\pm$0.13 \\
         ... &     ... &    \cf\ & $-$7.5$\pm$1.9 &  8.0$\pm$2.8 & 11.56$\pm$0.10 \\
         ... &     ... &    \hi\ &    2.2$\pm$1.1 & 37.2$\pm$1.1 & 13.73$\pm$0.01 \\
HE 1158-1843 &  2.4279 &    \os\ & $-$52.6$\pm$23.9 & 33.7$\pm$14 & 13.47$\pm$0.40 \\
         ... &     ... &     ... & $-$41.3$\pm$1.7 &  9.3$\pm$3.7 & 13.15$\pm$0.34 \\
         ... &     ... &     ... & $-$11.0$\pm$4.6 & 14.7$\pm$3.8 & 13.92$\pm$0.24 \\
         ... &     ... &     ... &    7.3$\pm$3.8 & 13.6$\pm$2.2 & 13.90$\pm$0.21 \\
         ... &     ... &    \cf\ & $-$5.4$\pm$1.2 & 20.7$\pm$1.4 & 12.71$\pm$0.02 \\
         ... &     ... &    \nf\ & $-$4.1$\pm$1.1 & 18.0$\pm$1.2 & 13.36$\pm$0.02 \\
         ... &     ... &    \hi\ & $-$8.4$\pm$1.0 & 32.8$\pm$1.1 & 13.71$\pm$0.01 \\
         ... &  2.4542 &    \os\ &    0.5$\pm$1.0 & 10.8$\pm$1.0 & 13.57$\pm$0.01 \\
         ... &     ... &     ... &    34.3$\pm$1.0 & 10.4$\pm$1.1 & 13.51$\pm$0.02 \\
         ... &     ... &     ... &    64.8$\pm$1.9 & 13.4$\pm$2.7 & 13.12$\pm$0.08 \\
         ... &     ... &     ... &    91.2$\pm$2.5 & 12.2$\pm$3.0 & 12.94$\pm$0.10 \\
         ... &     ... &    \cf\ &    31.4$\pm$1.4 &  8.0$\pm$1.9 & 11.92$\pm$0.06 \\
         ... &     ... &    \hi\ &    14.0$\pm$1.6 & 36.5$\pm$1.4 & 14.00$\pm$0.02 \\
         ... &     ... &     ... &    81.2$\pm$1.5 & 36.2$\pm$1.7 & 13.99$\pm$0.02 \\
HE 1341-1020 &  2.0850 &    \os\ &    2.4$\pm$1.1 & 12.1$\pm$1.3 & 13.97$\pm$0.02 \\
         ... &     ... &    \cf\ & $-$2.4$\pm$1.0 & 10.5$\pm$1.0 & 12.89$\pm$0.01 \\
         ... &     ... &    \hi\ & $-$82.6$\pm$1.3 & 23.8$\pm$1.2 & 13.77$\pm$0.02 \\
         ... &     ... &     ... & $-$3.4$\pm$1.1 & 37.0$\pm$1.1 & 15.05$\pm$0.02 \\
HE 1347-2457 &  2.5727 &    \os\ &    1.0$\pm$1.2 &  7.8$\pm$1.4 & 13.19$\pm$0.04 \\
         ... &     ... &    \cf\ &    1.6$\pm$1.1 & 10.6$\pm$1.3 & 12.44$\pm$0.03 \\
         ... &     ... &    \hi\ &    12.2$\pm$1.0 & 46.1$\pm$1.2 & 15.46$\pm$0.04 \\
         ... &  2.5744 &    \os\ &    7.4$\pm$1.2 & 22.0$\pm$1.3 & 13.45$\pm$0.02 \\
         ... &     ... &    \cf\ &    1.2$\pm$1.0 &  4.7$\pm$1.0 & 13.20$\pm$0.01 \\
         ... &     ... &     ... &    15.4$\pm$1.0 &  6.0$\pm$1.0 & 13.14$\pm$0.01 \\
         ... &     ... &    \hi\ &    8.7$\pm$1.0 & 15.8$\pm$1.0 & 14.95$\pm$0.02 \\
HE 2347-4342 &  2.8625 &    \os\ &    3.6$\pm$1.2 & 16.8$\pm$1.3 & 13.58$\pm$0.02 \\
         ... &     ... &    \cf\ & $-$7.6$\pm$1.0 &  9.7$\pm$1.1 & 12.79$\pm$0.02 \\
         ... &     ... &     ... &    13.2$\pm$1.1 &  7.2$\pm$1.2 & 12.45$\pm$0.03 \\
         ... &     ... &    \hi\ & $-$31.7$\pm$15.8 & 33.5$\pm$6.0 & 13.83$\pm$0.37 \\
         ... &     ... &     ... & $-$1.4$\pm$4.4 & 29.7$\pm$1.9 & 14.18$\pm$0.17 \\
\hline
\end{tabular}
\end{minipage}
\end{table}
\begin{table}
\centering
\scriptsize
\begin{minipage}{80mm}
\caption{Component fit results to weak \os\ absorbers (cont.)}
\begin{tabular}{lcccc c}
\hline
QSO & $z_{\rm abs}$ & Ion & $v_0$ (\kms) & $b$ (\kms) & log\,($N$ in \sqcm) \\
\hline
 PKS 0237-23 &  2.2028 &    \os\ &    4.3$\pm$1.1 & 26.5$\pm$1.4 & 13.80$\pm$0.02 \\
         ... &     ... &    \cf\ & $-$52.7$\pm$5.4 & 19.4$\pm$6.7 & 12.73$\pm$0.16 \\
         ... &     ... &     ... & $-$29.2$\pm$2.6 & 10.0$\pm$5.7 & 12.40$\pm$0.36 \\
         ... &     ... &     ... & $-$2.2$\pm$1.0 & 10.7$\pm$1.0 & 14.03$\pm$0.01 \\
         ... &     ... &     ... &    29.1$\pm$1.0 & 13.2$\pm$1.1 & 13.56$\pm$0.01 \\
         ... &     ... &    \nf\ & $-$2.1$\pm$1.6 &  7.2$\pm$2.6 & 11.81$\pm$0.08 \\
         ... &     ... &     ... &    27.2$\pm$2.9 &  6.6$\pm$4.5 & 11.52$\pm$0.18 \\
         ... &     ... &    \hi\ & $-$10.6$\pm$1.0 & 50.3$\pm$1.1 & 15.40$\pm$0.02 \\
         ... &  2.2135 &    \os\ &    2.6$\pm$1.0 &  8.9$\pm$1.1 & 13.40$\pm$0.02 \\
         ... &     ... &    \cf\ &    7.7$\pm$1.3 & 16.2$\pm$1.6 & 12.10$\pm$0.03 \\
         ... &     ... &    \hi\ &    11.3$\pm$1.0 & 26.7$\pm$1.0 & 14.25$\pm$0.01 \\
         ... &     ... &     ... &    73.9$\pm$1.4 & 24.4$\pm$1.8 & 12.77$\pm$0.03 \\
         ... &  2.2298 &    \os\ & $-$0.2$\pm$1.1 & 18.0$\pm$1.3 & 13.61$\pm$0.02 \\
         ... &     ... &     ... &    29.4$\pm$1.7 &  8.9$\pm$2.8 & 12.69$\pm$0.11 \\
         ... &     ... &    \cf\ & $-$0.9$\pm$1.0 &  7.5$\pm$1.1 & 12.22$\pm$0.02 \\
         ... &     ... &    \hi\ & $-$6.8$\pm$1.6 & 31.2$\pm$2.3 & 12.95$\pm$0.03 \\
         ... &  2.2363 &    \os\ &    6.8$\pm$1.1 & 15.2$\pm$1.1 & 14.26$\pm$0.02 \\
         ... &     ... &     ... &    39.7$\pm$1.6 &  6.8$\pm$2.6 & 13.32$\pm$0.09 \\
         ... &     ... &     ... &    62.5$\pm$3.0 &  7.3$\pm$5.8 & 13.01$\pm$0.19 \\
         ... &     ... &    \cf\ &    4.0$\pm$1.1 & 12.3$\pm$1.1 & 12.53$\pm$0.02 \\
         ... &     ... &     ... &    36.3$\pm$1.8 & 13.9$\pm$2.7 & 11.96$\pm$0.06 \\
         ... &     ... &    \nf\ &    2.7$\pm$1.9 & 11.1$\pm$2.6 & 11.96$\pm$0.07 \\
         ... &     ... &    \hi\ &    1.9$\pm$1.4 & 21.5$\pm$1.1 & 14.15$\pm$0.04 \\
         ... &     ... &     ... &    33.8$\pm$2.7 & 24.1$\pm$1.8 & 13.84$\pm$0.06 \\
         ... &  2.2378 &    \os\ & $-$3.7$\pm$1.0 &  8.3$\pm$1.0 & 13.56$\pm$0.02 \\
         ... &     ... &    \hi\ & $-$10.0$\pm$1.1 & 18.8$\pm$1.3 & 12.24$\pm$0.02 \\
PKS 0329-255 &  2.6494 &    \os\ &    0.0$\pm$1.1 &  8.6$\pm$1.2 & 13.14$\pm$0.02 \\
         ... &     ... &    \cf\ & $-$2.7$\pm$1.1 &  7.6$\pm$1.1 & 12.34$\pm$0.02 \\
         ... &     ... &    \hi\ & $-$5.6$\pm$1.0 & 25.5$\pm$1.0 & 14.67$\pm$0.01 \\
         ... &  2.6610 &    \os\ & $-$3.3$\pm$1.1 & 11.9$\pm$1.2 & 13.15$\pm$0.02 \\
         ... &     ... &    \hi\ & $-$7.8$\pm$1.0 & 26.2$\pm$1.0 & 14.30$\pm$0.01 \\
PKS 1448-232 &  2.1660 &    \os\ & $-$75.8$\pm$1.3 & 10.9$\pm$2.3 & 13.15$\pm$0.05 \\
         ... &     ... &     ... & $-$53.7$\pm$1.2 &  4.6$\pm$1.5 & 13.02$\pm$0.08 \\
         ... &     ... &     ... & $-$28.2$\pm$1.5 & 13.7$\pm$2.0 & 13.67$\pm$0.06 \\
         ... &     ... &     ... &    3.7$\pm$3.4 & 16.3$\pm$5.5 & 13.55$\pm$0.24 \\
         ... &     ... &     ... &    34.9$\pm$10.8 & 22.8$\pm$11 & 13.43$\pm$0.27 \\
         ... &     ... &    \cf\ & $-$30.5$\pm$1.4 &  9.0$\pm$2.2 & 11.99$\pm$0.06 \\
         ... &     ... &     ... &    0.7$\pm$1.3 &  9.5$\pm$1.1 & 13.15$\pm$0.05 \\
         ... &     ... &     ... &    16.7$\pm$1.6 & 10.6$\pm$2.3 & 12.95$\pm$0.11 \\
         ... &     ... &     ... &    39.1$\pm$2.2 & 11.6$\pm$2.2 & 12.47$\pm$0.09 \\
         ... &     ... &     ... &    144.3$\pm$1.7 & 21.2$\pm$2.3 & 12.34$\pm$0.03 \\
         ... &     ... &     ... &    186.4$\pm$1.3 &  6.5$\pm$1.6 & 11.93$\pm$0.05 \\
         ... &     ... &    \nf\ &     9.6$\pm$2.4 &  9.2$\pm$3.7 & 12.02$\pm$0.12 \\
         ... &     ... &    \hi\ & $-$16.1$\pm$ 9.6 & 46.0$\pm$2.0 & 14.81$\pm$0.34 \\
         ... &     ... &     ... &    1.3$\pm$3.6 & 37.5$\pm$3.3 & 15.15$\pm$0.15 \\
 Q 0109-3518 &  2.4012 &    \os\ & $-$3.1$\pm$1.1 &  7.1$\pm$1.3 & 12.81$\pm$0.04 \\
         ... &     ... &    \cf\ & $-$3.5$\pm$2.3 &  8.4$\pm$3.0 & 11.48$\pm$0.11 \\
         ... &     ... &    \hi\ & $-$46.7$\pm$2.1 & 25.3$\pm$1.7 & 12.80$\pm$0.05 \\
         ... &     ... &     ... &    0.5$\pm$1.0 & 28.1$\pm$1.1 & 13.77$\pm$0.01 \\
         ... &     ... &     ... &    45.1$\pm$1.3 & 20.3$\pm$1.2 & 12.85$\pm$0.03 \\
  Q 0122-380 &  2.1587 &    \os\ &    16.1$\pm$1.6 &  9.4$\pm$2.5 & 13.18$\pm$0.08 \\
         ... &     ... &    \cf\ & $-$18.2$\pm$1.1 &  9.6$\pm$1.2 & 12.48$\pm$0.03 \\
         ... &     ... &     ... &    4.7$\pm$1.0 &  9.9$\pm$1.0 & 13.35$\pm$0.01 \\
         ... &     ... &    \hi\ & $-$73.5$\pm$3.1 & 31.7$\pm$1.8 & 14.18$\pm$0.11 \\
         ... &     ... &     ... &    0.4$\pm$1.1 & 24.1$\pm$1.6 & 16.36$\pm$0.15 \\
         ... &     ... &     ... &    16.2$\pm$20.0 & 57.7$\pm$7.5 & 14.48$\pm$0.26 \\
  Q 0329-385 &  2.3520 &    \os\ &    7.3$\pm$1.1 & 12.6$\pm$1.1 & 14.06$\pm$0.02 \\
         ... &     ... &    \cf\ &    1.1$\pm$1.1 &  6.4$\pm$1.0 & 13.31$\pm$0.04 \\
         ... &     ... &     ... &    11.9$\pm$1.2 &  6.4$\pm$1.2 & 13.02$\pm$0.06 \\
         ... &     ... &    \nf\ &    4.2$\pm$1.0 & 10.2$\pm$1.1 & 13.60$\pm$0.02 \\
         ... &     ... &    \hi\ &    13.6$\pm$1.3 & 20.6$\pm$1.6 & 13.05$\pm$0.02 \\
         ... &  2.3639 &    \os\ & $-$46.7$\pm$1.2 & 15.6$\pm$1.5 & 13.38$\pm$0.03 \\
         ... &     ... &     ... & $-$3.6$\pm$1.1 & 11.0$\pm$1.2 & 13.55$\pm$0.02 \\
         ... &     ... &    \cf\ & $-$50.3$\pm$1.7 &  8.1$\pm$2.4 & 11.86$\pm$0.08 \\
         ... &     ... &     ... & $-$10.4$\pm$1.2 &  8.9$\pm$1.4 & 12.24$\pm$0.03 \\
         ... &     ... &    \hi\ & $-$29.9$\pm$1.0 & 32.3$\pm$1.0 & 14.95$\pm$0.02 \\
         ... &  2.3738 &    \os\ & $-$118.9$\pm$26.1 & 45.4$\pm$18 & 13.54$\pm$0.32 \\
         ... &     ... &     ... & $-$74.3$\pm$3.1 & 27.2$\pm$5.4 & 13.73$\pm$0.21 \\
         ... &     ... &     ... & $-$11.7$\pm$1.3 & 26.2$\pm$1.6 & 13.94$\pm$0.02 \\
         ... &     ... &    \cf\ & $-$84.3$\pm$2.0 & 22.5$\pm$2.8 & 12.42$\pm$0.04 \\
         ... &     ... &     ... & $-$25.5$\pm$1.4 & 16.2$\pm$1.9 & 12.46$\pm$0.03 \\
         ... &     ... &    \nf\ & $-$5.3$\pm$1.9 &  6.2$\pm$3.3 & 11.94$\pm$0.12 \\
         ... &     ... &    \hi\ & $-$151.8$\pm$3.4 & 46.0$\pm$2.1 & 13.96$\pm$0.05 \\
         ... &     ... &     ... & $-$53.8$\pm$1.6 & 50.6$\pm$1.4 & 15.15$\pm$0.03 \\
         ... &     ... &     ... & $-$26.3$\pm$3.4 & 26.1$\pm$3.4 & 14.53$\pm$0.11 \\
  Q 0453-423 &  2.6362 &    \os\ &    3.3$\pm$1.1 & 15.5$\pm$1.1 & 13.76$\pm$0.02 \\
         ... &     ... &    \cf\ & $-$1.8$\pm$1.0 &  9.7$\pm$1.0 & 13.64$\pm$0.01 \\
         ... &     ... &    \nf\ & $-$2.4$\pm$1.2 & 11.9$\pm$1.4 & 12.41$\pm$0.03 \\
         ... &     ... &    \hi\ &    1.4$\pm$1.2 & 21.1$\pm$1.6 & 14.80$\pm$0.02 \\
         ... &  2.6402 &    \os\ &    4.1$\pm$1.0 &  6.3$\pm$1.0 & 13.16$\pm$0.02 \\
         ... &     ... &    \cf\ &    0.0$\pm$1.0 &  7.2$\pm$1.0 & 12.43$\pm$0.01 \\
         ... &     ... &    \nf\ & $-$3.9$\pm$1.9 &  4.8$\pm$3.1 & 11.64$\pm$0.14 \\
         ... &     ... &    \hi\ & $-$0.6$\pm$1.0 & 25.5$\pm$1.1 & 14.66$\pm$0.01 \\
\hline
\end{tabular}
\end{minipage}
\end{table}

\clearpage

% FIGURES
\setcounter{figure}{0}
\begin{figure*}
\includegraphics[width=185mm]{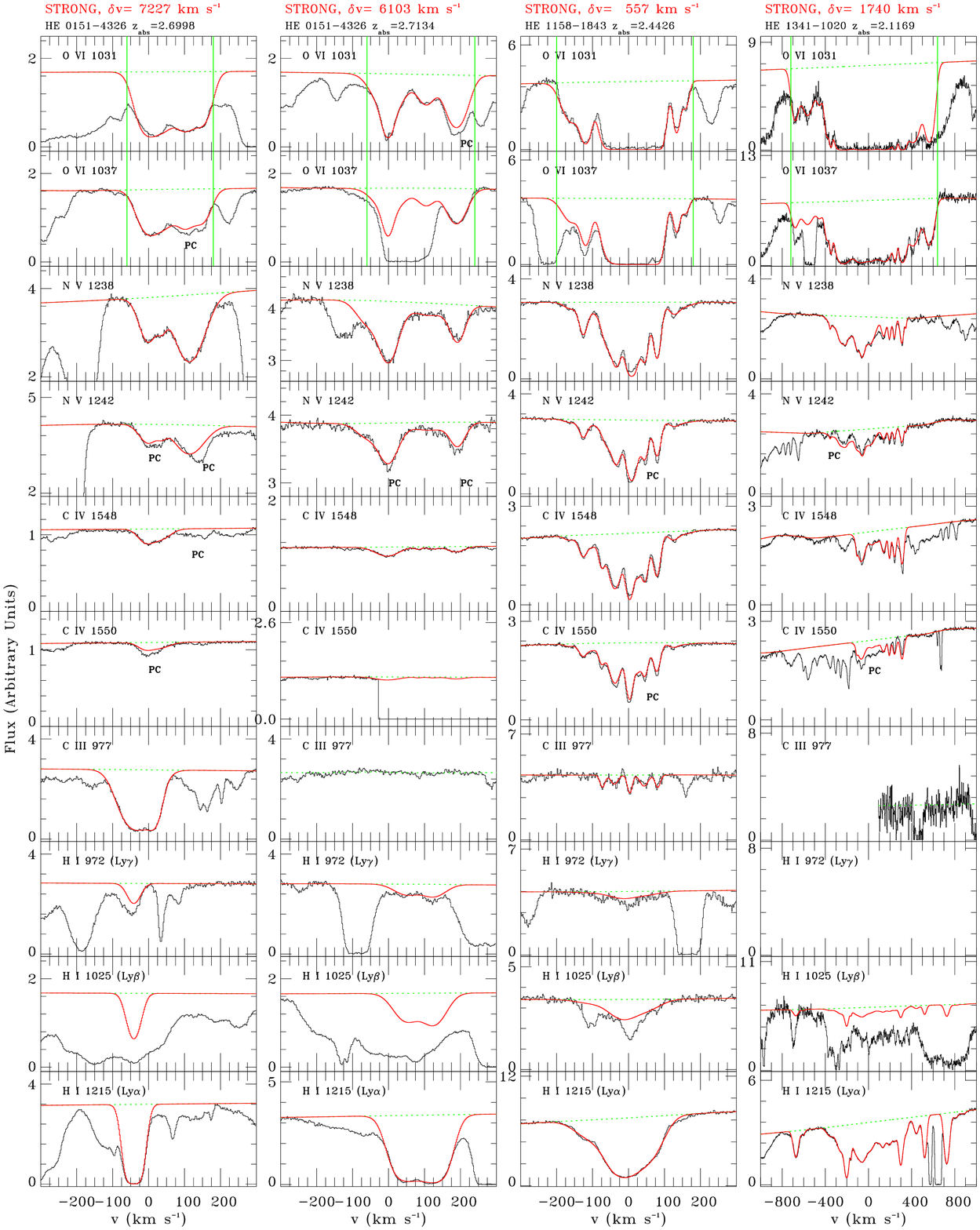}
\caption{Absorption-line spectra of all proximate \os\ systems in the
 UVES Large Programme. 
 Strong absorbers are shown first, followed by the weak absorbers.
 Our Voigt profile fits are shown in red. 
 In many of the \cf\ and \nf\ panels, the data
 are shown over a limited range in the y-axis, for clarity.
 Green vertical lines indicate the velocity range of the AOD integration.
 The label `PC' indicates regions where partial coverage of the
 continuum source affects the line profiles.
 At the top of each column we annotate our classification and the
 proximity of the absorber to the quasar.} 
\end{figure*}
\addtocounter{figure}{-1}
\begin{figure*}
\includegraphics[width=195mm]{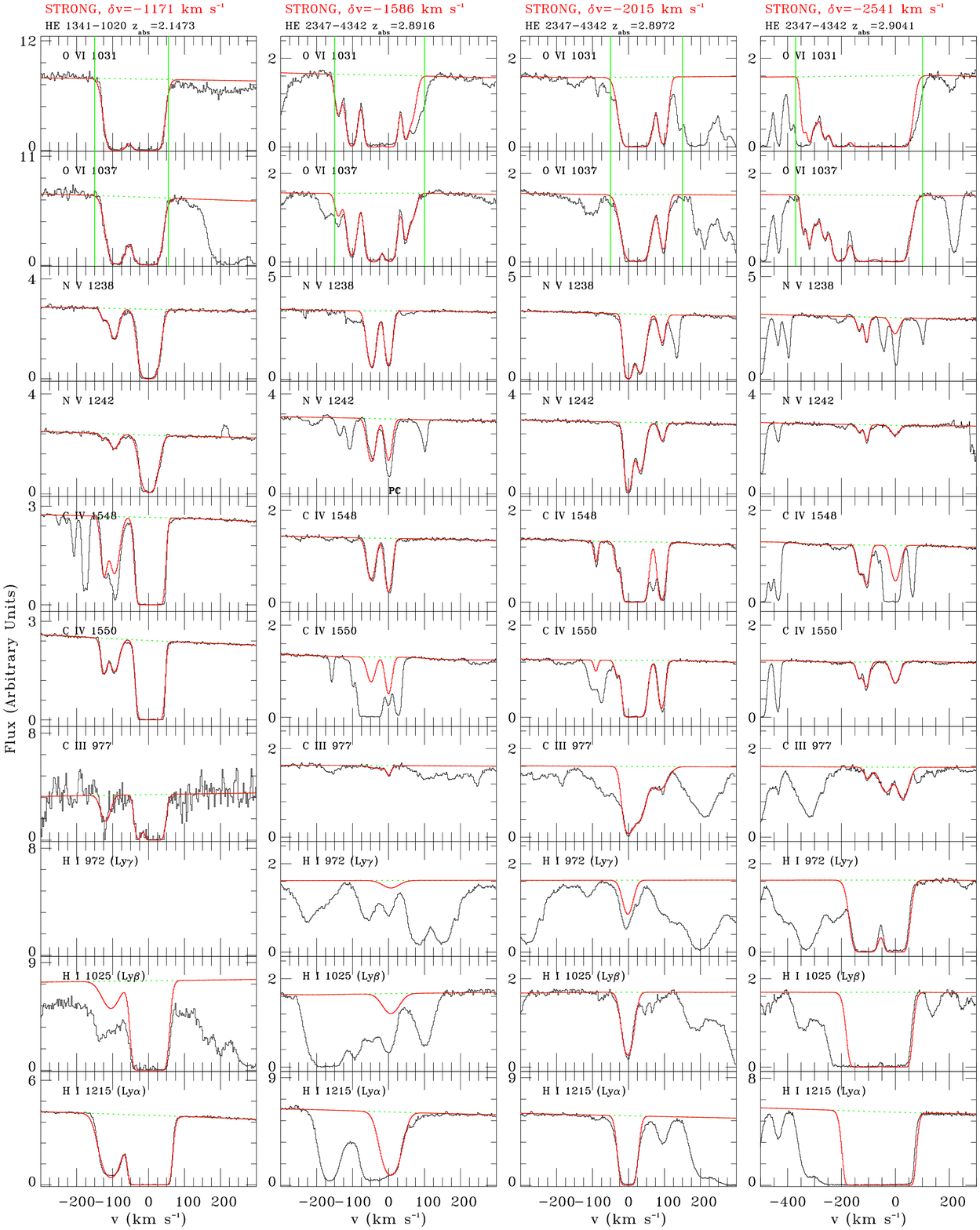}
\caption{(cont.) }
\end{figure*}
\addtocounter{figure}{-1}
\begin{figure*}
\includegraphics[width=195mm]{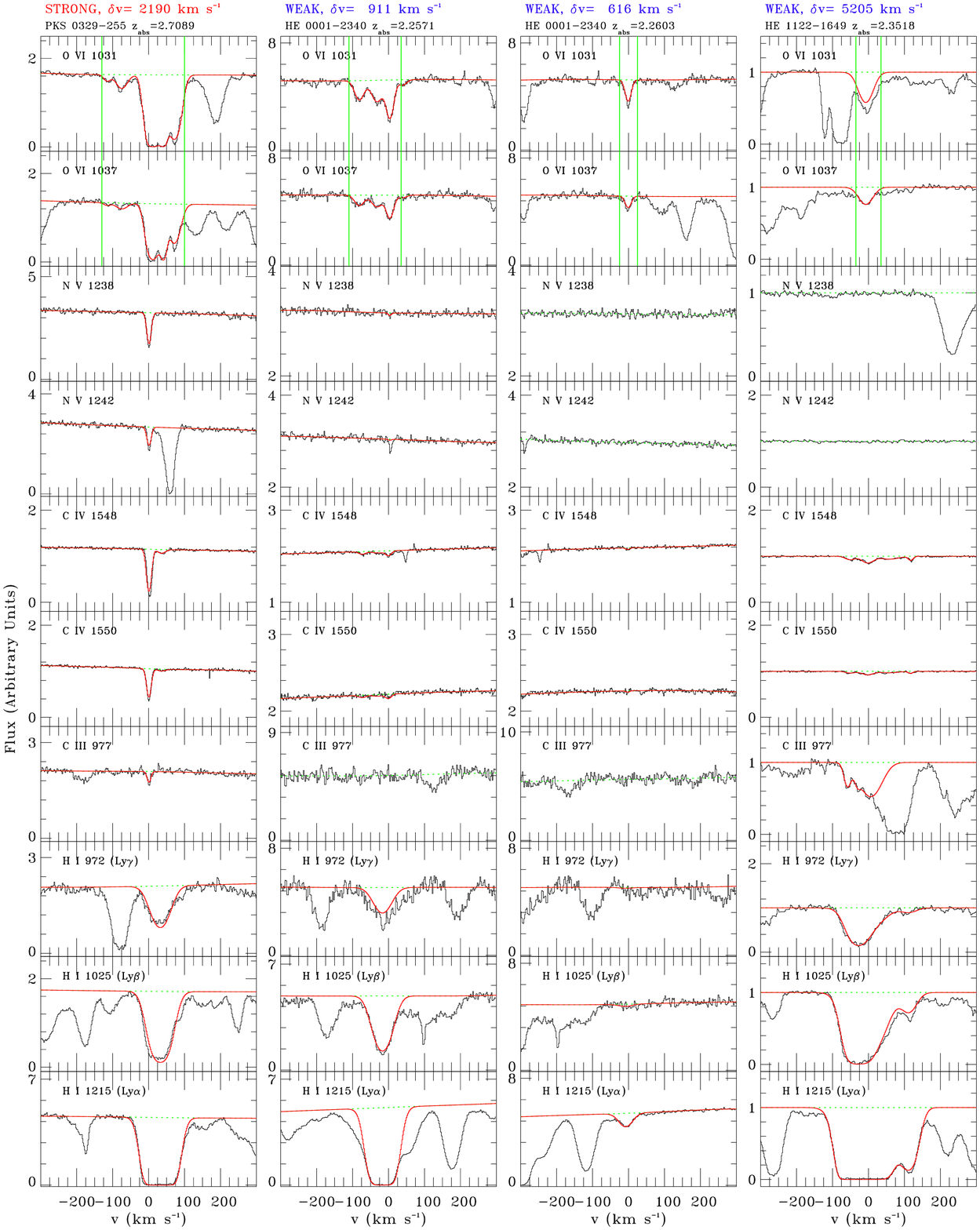}
\caption{(cont.) }
\end{figure*}

\addtocounter{figure}{-1}
\begin{figure*}
\includegraphics[width=195mm]{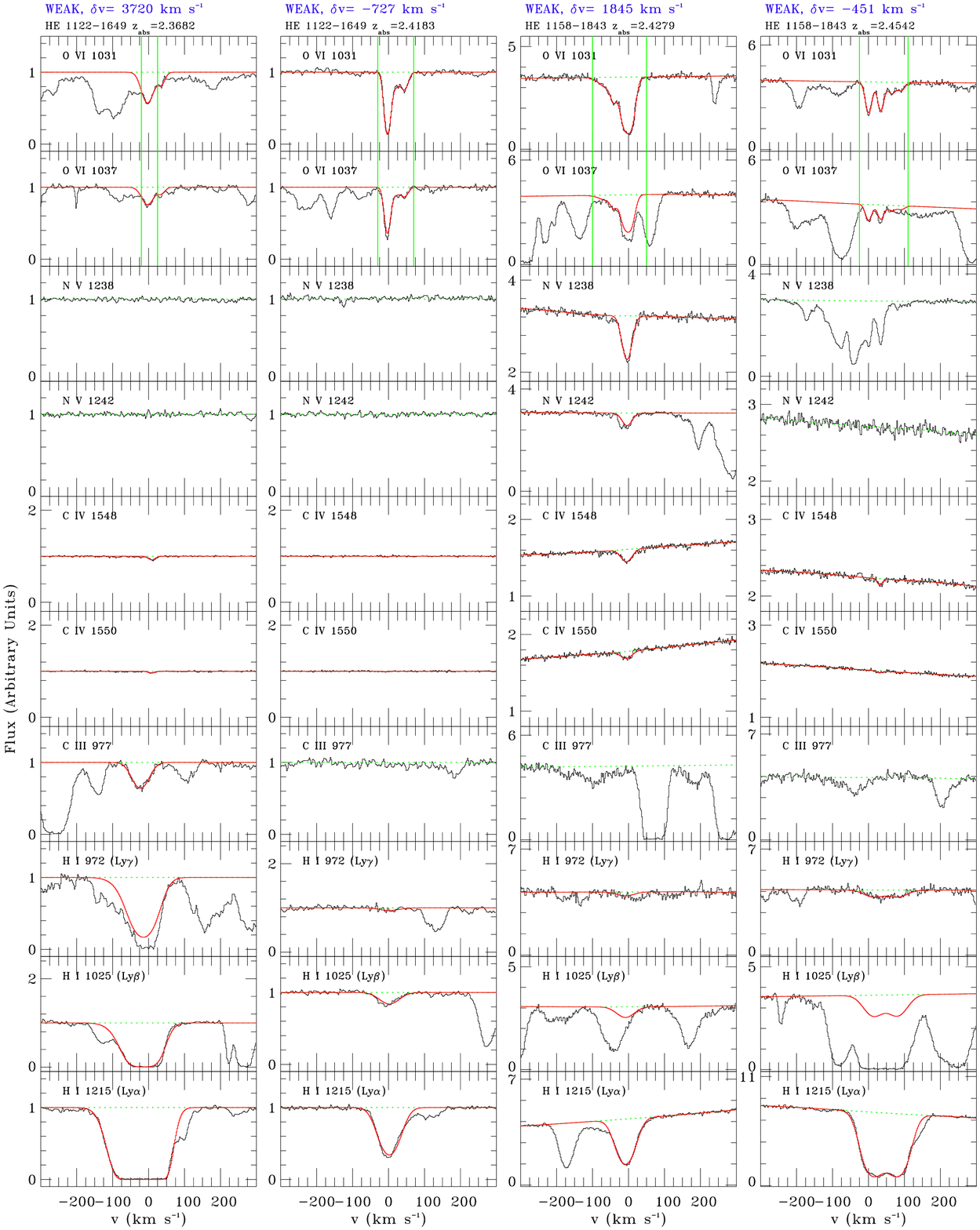}
\caption{(cont.) }
\end{figure*}
\addtocounter{figure}{-1}
\begin{figure*}
\includegraphics[width=195mm]{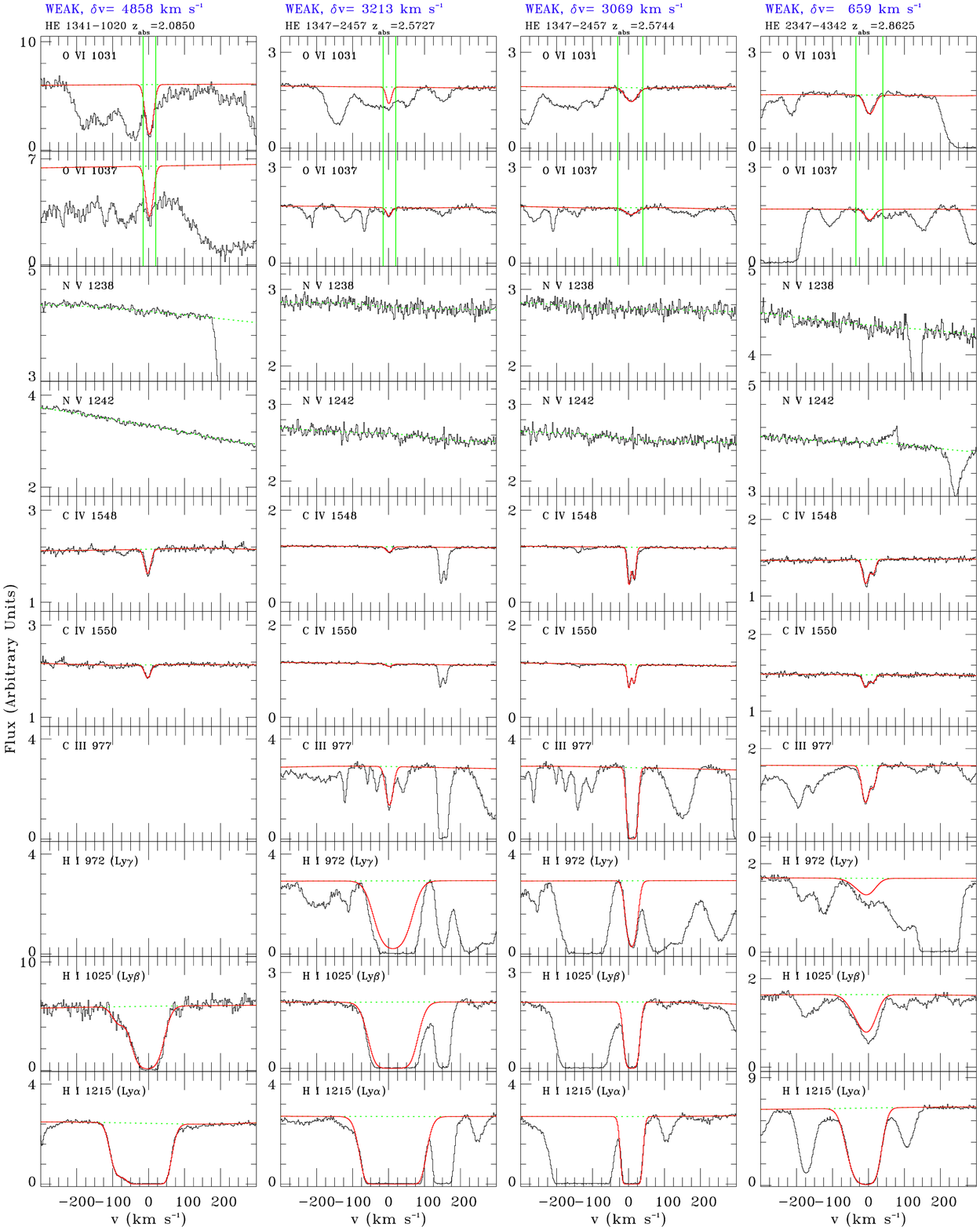}
\caption{(cont.) }
\end{figure*}
\addtocounter{figure}{-1}
\begin{figure*}
\includegraphics[width=195mm]{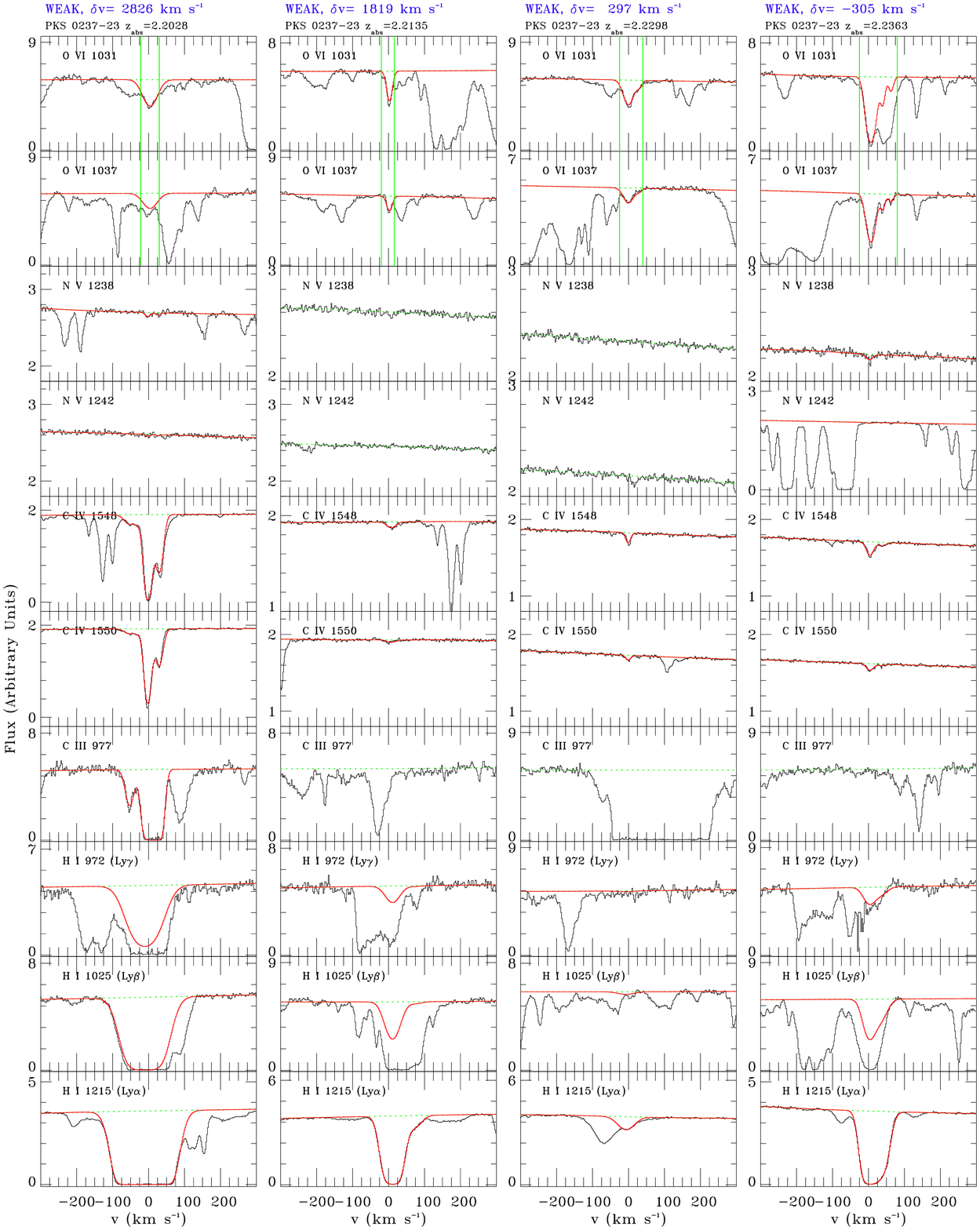}
\caption{(cont.) }
\end{figure*}

\addtocounter{figure}{-1}
\begin{figure*}
\includegraphics[width=195mm]{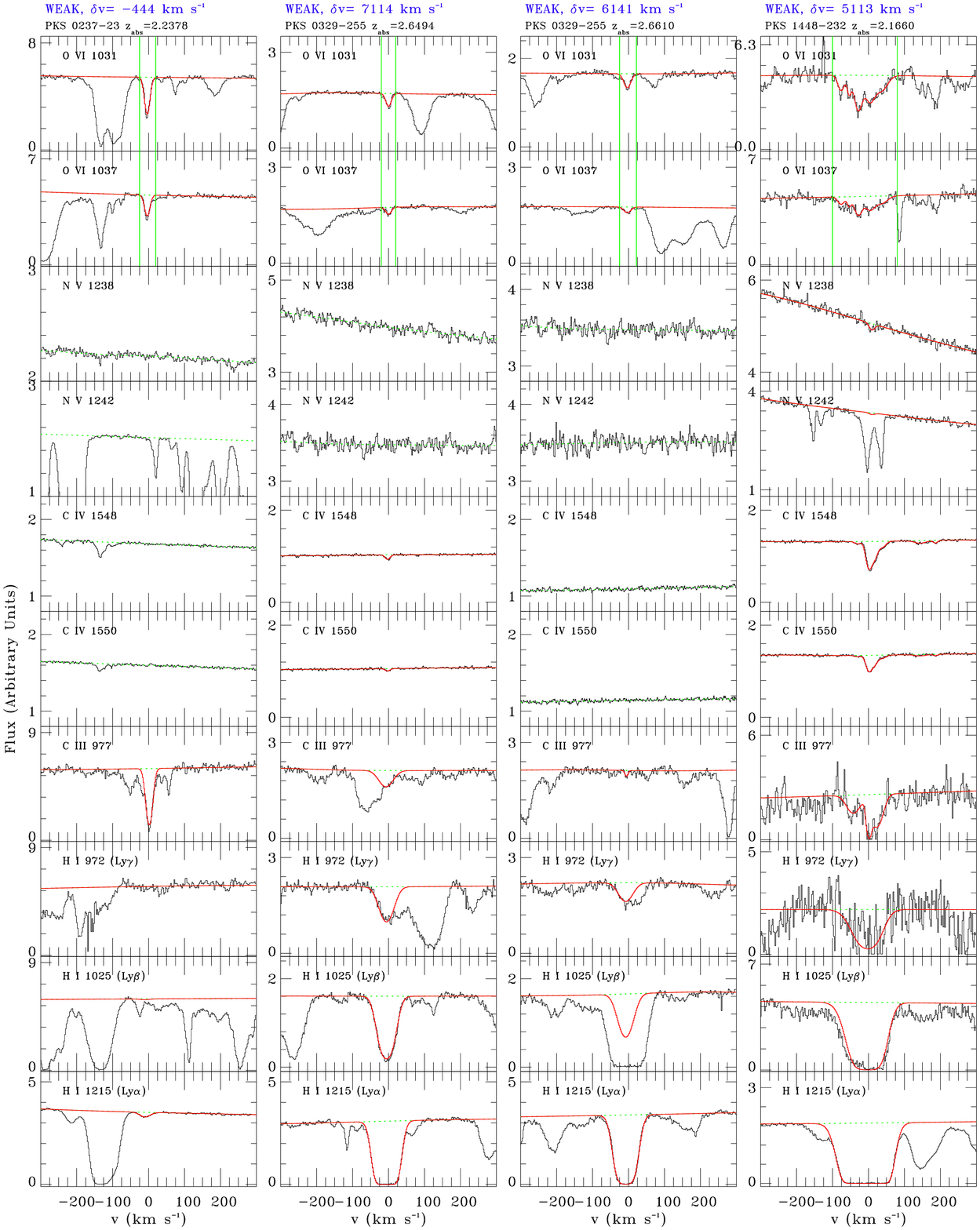}
\caption{(cont.) }
\end{figure*}

\addtocounter{figure}{-1}
\begin{figure*}
\includegraphics[width=195mm]{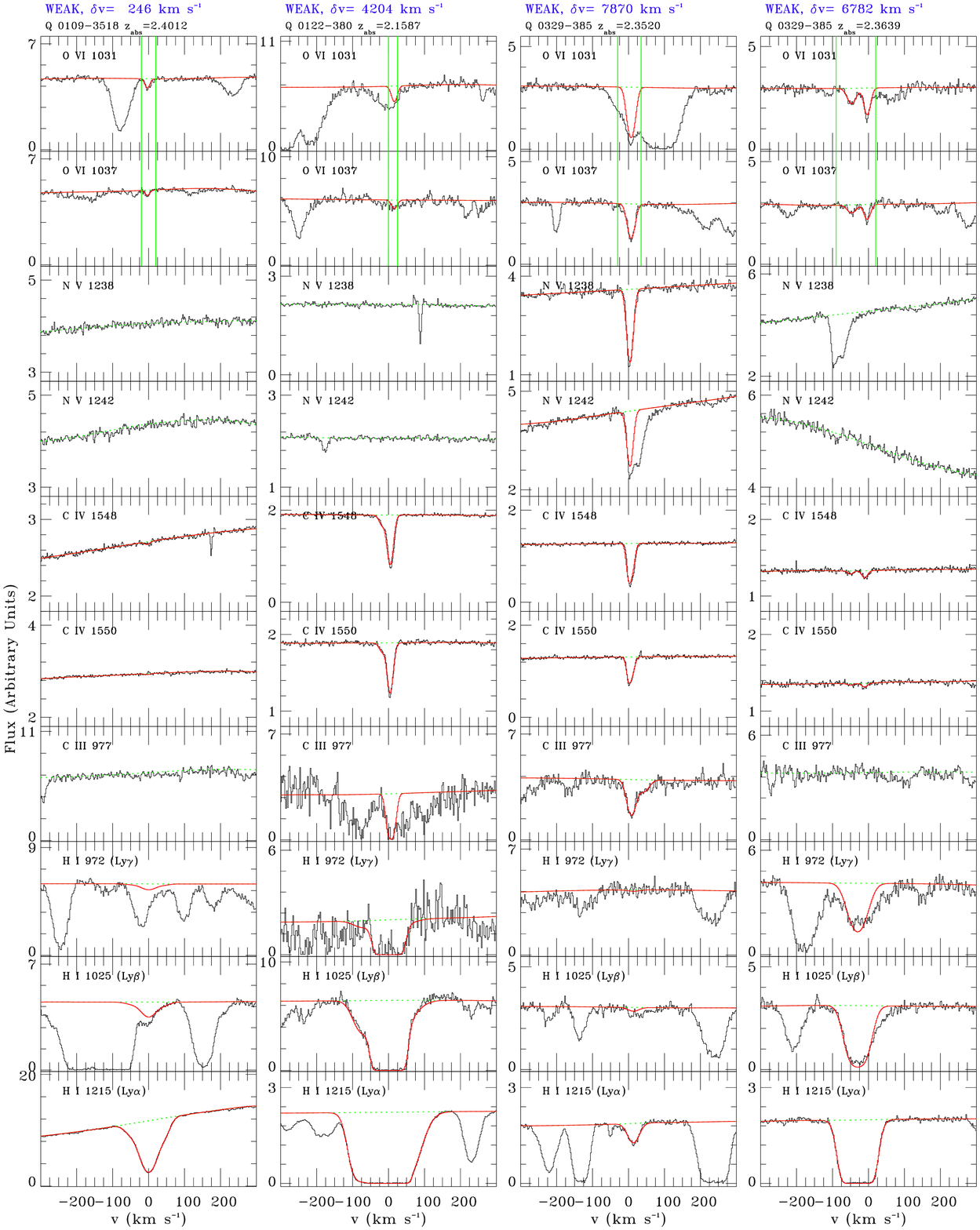}
\caption{(cont.) }
\end{figure*}
\addtocounter{figure}{-1}
\begin{figure*}
\includegraphics[width=195mm]{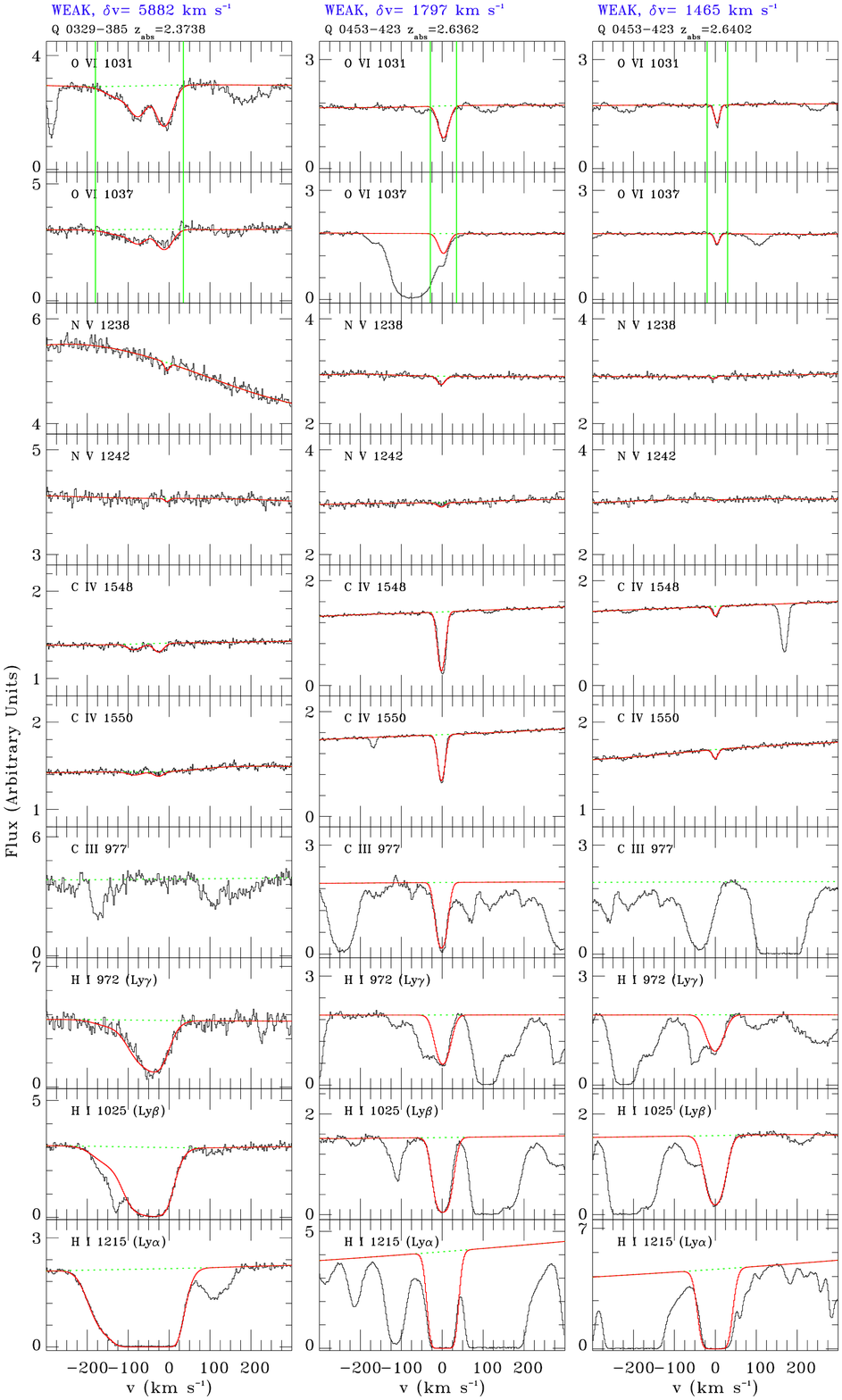}
\caption{(cont.) }
\end{figure*}

\end{document}